\documentclass[aps,prx,twocolumn,superscriptaddress,nofootinbib,floatfix]{revtex4-2}

\usepackage{amsmath,amssymb,amsfonts}
\usepackage{graphicx}
\usepackage{hyperref}
\usepackage{xcolor}
\usepackage{algorithm}
\usepackage{algorithmic}
\usepackage{booktabs}
\usepackage{enumitem}
\usepackage{multirow}
\usepackage{bm}
\usepackage{braket}
\usepackage[normalem]{ulem}
\usepackage{tikz}
\usetikzlibrary{arrows.meta, positioning, matrix}
\newcommand{\R}{\mathbb{R}}

\newcommand{\etal}{\textit{et al.}}

\def \gp #1 {\textcolor{blue}{#1}}
\def \hjz #1 {\textcolor{brown}{#1}}
\def \zgq #1 {\textcolor{cyan}{#1}}

\begin{document}

\title{QuantiSpect: A Structure-Aware Lightweight 3D CNN Pre-Decoder for\\Scalable Surface Code Quantum Error Correction}

\author{Pan Gao}
\thanks{These authors contributed equally to this work.}
\affiliation{Beijing Academy of Quantum Information Sciences, Beijing 100193, China}

\author{Xu-Sheng Xu}
\thanks{These authors contributed equally to this work.}
\affiliation{QuSpect Technology Co., Ltd., Beijing 100084, China}

\author{Ji-Ze Han}
\thanks{These authors contributed equally to this work.}
\affiliation{China Mobile (Suzhou) Software Technology Co., Ltd., Suzhou 215163, China}

\author{Jing-Wei Wen}
\affiliation{China Mobile (Suzhou) Software Technology Co., Ltd., Suzhou 215163, China}

\author{Ling Qian}
\affiliation{China Mobile (Suzhou) Software Technology Co., Ltd., Suzhou 215163, China}

\author{Xudong Lv}
\affiliation{Shanghai Institute of Optics and Fine Mechanics, Chinese Academy of Sciences, Shanghai 201800, China}

\author{Run-Qing Zhang}
\email[Corresponding author: ]{zrq1993@bupt.cn}
\affiliation{China Mobile (Suzhou) Software Technology Co., Ltd., Suzhou 215163, China}

\author{Xiao-Xiao Hu}
\email[Corresponding author: ]{xx-hu15@tsinghua.org.cn}
\affiliation{Shanghai Engine Fund, Shanghai 200010, China}

\author{Gui-Lu Long}
\email[Lead Corresponding author: ]{gllong@tsinghua.edu.cn}
\affiliation{Beijing Academy of Quantum Information Sciences, Beijing 100193, China}
\affiliation{State Key Laboratory of Low-Dimensional Quantum Physics and Department of Physics, Tsinghua University, Beijing 100084, China}
\affiliation{Frontier Science Center for Quantum Information, Beijing 100084, China}
\affiliation{Beijing National Research Center for Information Science and Technology, Beijing 100084, China}

\date{\today}

\begin{abstract}
Real-time decoding is a critical bottleneck for large-scale fault-tolerant quantum computing.
AI-based neural pre-decoders locally correct the majority of physical errors before passing residual syndromes to a global decoder, potentially enabling sub-microsecond decoding latencies.
However, existing architectures carry significant computational overhead due to their reliance on dense three-dimensional convolutions.
We present \textsc{QuantiSpect}, a lightweight 3D convolutional neural network (CNN) pre-decoder for the rotated surface code, built upon the decoding pipeline of Chamberland~\etal~\cite{chamberland2026fast}.
The key idea is to replace the dense 3D convolutions with three parallel branches inside each residual block: a depthwise spatial branch, a depthwise temporal branch, and a grouped spatio-temporal branch, followed by a squeeze-and-excitation channel gate.
This design reflects the known structure of surface code errors, where spatial lattice patterns and temporal syndrome correlations are partially separable.
On a unified 4$\times$A100 GPU benchmark, \textsc{QuantiSpect} matches the receptive field of the \textsc{Accurate} baseline with $R=13$ while using $\sim2.71\times$ fewer parameters ($0.663$\,M vs.\ $1.80$\,M) and $\sim2.84\times$ fewer per-voxel convolutional MACs ($0.633$\,M vs.\ $1.797$\,M).
Despite this reduction, it matches \textsc{Accurate}'s circuit-level threshold and its decoding accuracy at moderate and large code distances, reduces the logical error rate by up to $\sim1.85\times$ relative to uncorrelated PyMatching alone at the typical case $d=13$, $p=0.5\%$, and speeds up the PyMatching decode by up to $3.11\times$ relative to standalone PyMatching at a larger distance $d=23$.
We also explored enlarging the receptive field by increasing the number of blocks. Even at $R = 21$, the resulting model uses only $1.18$\,M parameters, fewer than the $R = 13$ \textsc{Accurate} baseline's $1.80$\,M and the $R=17$ dense model's $4.22$\,M of Ref.~\cite{chamberland2026fast} despite the larger receptive field.
This expanded variant performs significantly  better than the \textsc{Accurate} model in ~\cite{chamberland2026fast} by raising the circuit-level threshold to $p_{\text{th}} \approx 0.80\%$ and further reducing the logical error rate.
Together, both variants show that a structure-aware factorized design is an effective, parameter-efficient alternative to a dense one for decoding the surface code.
The model and code are publicly available at \url{https://huggingface.co/quantispect/QuantiSpect-V1}.
\end{abstract}

\maketitle

\section{Introduction}
\label{sec:introduction}

Quantum error correction (QEC) is an essential ingredient for scalable fault-tolerant quantum computing~\cite{shor1995scheme,steane1996error,gottesman1997stabilizer,knill1998resilient,terhal2015quantum}.
The surface code~\cite{kitaev2003fault,bravyi1998quantum,dennis2002topological} is a leading candidate for near-term implementations because it achieves a high threshold under circuit-level noise and is compatible with planar qubit layouts.
It also supports fault-tolerant logical operations through lattice surgery~\cite{horsman2012surface,litinski2019game,fowler2018low}.

Real-time decoding of the surface code poses significant computational challenges~\cite{battistel2023real}.
As the number of physical qubits scales and the error correction cycle frequency increases, the decoding process must be completed continuously within microsecond-level time budgets.
Conventional global decoders, including minimum-weight perfect matching (MWPM)~\cite{dennis2002topological,higgott2025sparse} and Union Find~\cite{delfosse2021almost}, achieve excellent logical error rate (LER) performance but exhibit syndrome-density-dependent runtimes.
In particular, MWPM scales as $\mathcal{O}(s^3)$ in the syndrome density $s$, making it a potential bottleneck at large code distances or near the error threshold.

\begin{figure}[ht]
\centering
\includegraphics[width=0.7\columnwidth]{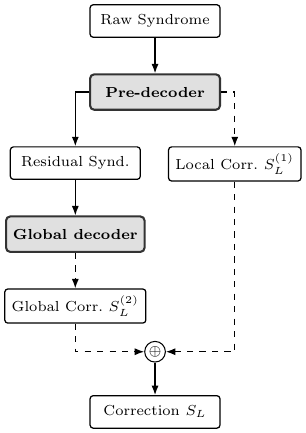}
\caption{Two-stage decoding workflow.}
\label{fig:two_stage_workflow}
\end{figure}

AI-based pre-decoders offer a promising approach to addressing this challenge~\cite{torlai2017neural,chamberland2026fast,bausch2024learning,meinerz2022scalable,gicev2023scalable,wang2023transformer,varsamopoulos2020decoding,ni2020neural}. Rather than replacing the global decoder, a fast neural pre-decoder corrects most errors locally, producing a substantially sparser residual syndrome that is subsequently processed by an algorithmic decoder capable of handling long-range correlations. This hybrid workflow, illustrated in Figure~\ref{fig:two_stage_workflow}, reduces the computational burden on the global decoder while preserving its ability to resolve nonlocal errors.

Recently, Chamberland~\etal~\cite{chamberland2026fast} introduced a fully convolutional 3D neural network pre-decoder for the rotated surface code, along with data processing techniques such as spacelike and timelike homological equivalence that improve training label quality.
With uncorrelated PyMatching~\cite{higgott2025sparse} as the global decoder, their pipeline reaches $\mathcal{O}(1\,\mu\text{s})$ per syndrome round on NVIDIA GB300 GPUs while reducing logical error rates.
The architecture uses a stack of dense 3D convolutional layers with $128$ or $256$ filters per layer, yielding models with $0.9$--$7.1$ million parameters.

However, the dense 3D convolutions carry significant computational overhead. Each hidden layer applies full cross-channel mixing at every voxel, so both the parameter count and the per-voxel MAC cost scale quadratically in the channel width. This design does not exploit the known property of surface code errors that spatial error patterns and temporal measurement correlations are partially separable.
A more efficient architecture would be valuable for low-latency, low-cost, or resource-limited settings, provided that decoding accuracy is preserved.

In this work, we introduce \textsc{QuantiSpect}, a lightweight 3D CNN pre-decoder that serves as a drop-in replacement for the dense backbone in Ref.~\cite{chamberland2026fast}.
Its main body is built from five identical FastHyperBlocks, each containing three factorized parallel branches, matching the receptive field of the \textsc{Accurate} baseline in Ref.~\cite{chamberland2026fast} with $R=13$ while cutting the parameters by $\sim2.71\times$ and the  per-voxel MACs by $\sim2.84\times$.
Under a unified benchmark on four A100 GPUs, \textsc{QuantiSpect} achieves the same circuit-level threshold as \textsc{Accurate}. Wilson-weighted finite-size scaling yields $p_{\text{th}}=0.769\%$ for both decoders. \textsc{QuantiSpect} also matches the decoding accuracy of \textsc{Accurate} at moderate and large code distances. Compared with uncorrelated PyMatching, \textsc{QuantiSpect} achieves up to a $\sim1.85\times$ reduction in logical error rate at $d=13$ and $p=0.5\%$, while delivering a decoding speedup of up to $3.11\times$ over standalone PyMatching at $d=23$.

The main idea is to replace the dense $3\times 3\times 3$ convolution with three parallel branches inside each residual block: a depthwise spatial branch ($1\times 3\times 3$), a depthwise temporal branch ($3\times 1\times 1$), and a grouped spatio-temporal branch ($3\times 3\times 3$).
This decomposition encodes the partial separability of spatial and temporal error patterns rather than learning it from scratch.
A squeeze-and-excitation gate after each block re-weights channels based on their learned importance.
Each block uses GroupNorm for better stability on sparse binary syndrome inputs, unlike the baseline's Conv3D stack, which has no normalization layer at all. Pre-activation residual connections~\cite{he2016identity} further improve gradient flow, while the depthwise factorization itself draws on efficient CNN design principles~\cite{howard2017mobilenets,sandler2018mobilenetv2}.

Apart from the network backbone itself, we retain the full data processing pipeline, training methodology, and residual syndrome construction from Ref.~\cite{chamberland2026fast}.
This means \textsc{QuantiSpect} is a drop-in replacement. Only the CNN architecture changes, and the rest of the decoding framework remains the same.

The modular block design follows a simple receptive-field formula, $R = 1 + 2 + 2N$. Each block costs only $\sim128$\,k parameters (Sec.~\ref{sec:mac}), far less than a single dense convolutional layer of comparable width. This lower per-block cost means the receptive field can be extended by adding blocks at a much smaller parameter cost than stacking more dense layers or using larger kernels would require.
We demonstrate this by constructing an $R=21$ variant with $N=9$ blocks, which uses only $1.18$\,M parameters. This is fewer than both the $R=13$ \textsc{Accurate} baseline's $1.80$\,M and the $R=17$ dense model's $4.22$\,M despite a larger receptive field.
This expanded variant is also more accurate than the $R=13$ version since it reduces the logical error rate by a further $31\%$ at $d=23$, and raises the circuit-level threshold to $p_{\text{th}} \approx 0.80\%$.
These two variants show that a structure-aware factorized design is an effective, parameter-efficient alternative to a dense one for decoding the surface code.

The remainder of this paper is organized as follows.
Section~\ref{sec:background} reviews the rotated surface code, the syndrome extraction process, and the pre-decoding paradigm.
Section~\ref{sec:method} describes the \textsc{QuantiSpect} architecture in detail, including the factorized block design, its parameter and computational cost, the data format, and the training procedure.
Section~\ref{sec:results} presents numerical results comparing \textsc{QuantiSpect}-13 with the \textsc{Fast} and \textsc{Accurate} baselines across code distances $d = 5$ to $23$, covering threshold estimation, logical error rates, decoding speedup, and effective code distance.
Section~\ref{sec:rf_scaling} studies receptive field scaling with the deeper \textsc{QuantiSpect}-21 variant and compares it with the dense models from Ref.~\cite{chamberland2026fast}.
Section~\ref{sec:conclusion} concludes with a summary and directions for future work.

\section{Background}
\label{sec:background}

\subsection{Rotated surface code and syndrome extraction}
\label{sec:surface_code}

The rotated surface code~\cite{kitaev2003fault,bravyi1998quantum,bombin2007optimal,dennis2002topological} encodes one logical qubit ($k=1$) into $d^2$ data qubits and $(d^2-1)$ ancilla qubits arranged on a two-dimensional lattice, where $d$ is the code distance.
It is optimal for a planar topological code with the standard notation $[\![d^2,1,d]\!]$~\cite{bombin2007optimal}.

\begin{figure}[ht]
\centering
\includegraphics[width=0.9\columnwidth]{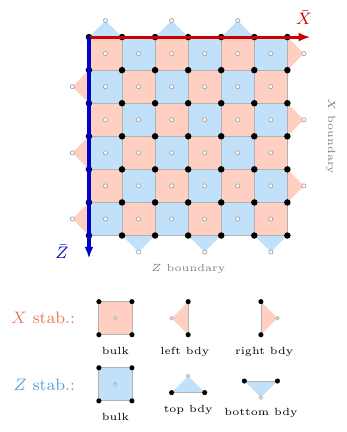}
\caption{Rotated surface code patch for $d = 7$.
The $d^2 = 49$ data qubits (black circles) reside on the vertices of a $7 \times 7$ square lattice.
$X$-type stabilizers (orange plaquettes) and $Z$-type stabilizers (blue plaquettes) alternate in a checkerboard pattern.
Bulk stabilizers are weight-four operators, boundary stabilizers (triangles at the edges) are weight-two operators.
Open circles mark ancilla qubit positions at plaquette centers.
Minimum-weight representatives of the logical operators $\bar{X}$ (red, horizontal) and $\bar{Z}$ (blue, vertical) connect opposite boundaries of the matching type.}
\label{fig:surface_code}
\end{figure}

Figure~\ref{fig:surface_code} shows a $d = 7$ example.
Data qubits (black circles) sit on the vertices of a $d \times d$ square lattice.
Ancilla qubits (open circles) sit at the centers of the plaquettes and are used to measure the stabilizer operators.
The $X$-type and $Z$-type stabilizers are arranged in a checkerboard pattern.
Each $X$-type stabilizer (orange) is a product of Pauli-$X$ operators on the data qubits at the corners of its plaquette, each $Z$-type stabilizer (blue) is a product of Pauli-$Z$ operators on the corresponding corners.

In the bulk, each plaquette touches four data qubits, giving weight-four stabilizers.
At the boundary, plaquettes are cut into triangles that touch only two data qubits, giving weight-two stabilizers.
The left and right boundaries carry weight-two $X$-type stabilizers (orange triangles). The top and bottom boundaries carry weight-two $Z$-type stabilizers (blue triangles).
In total there are $d^2 - 1$ stabilizer generators, split evenly as $K_X = K_Z = (d^2-1)/2$.

The logical operators $\bar{X}$ and $\bar{Z}$ are strings of Pauli operators that cross the full lattice between opposite boundaries of the matching type.
$\bar{X}$ runs horizontally (left to right, $d$ Pauli-$X$ operators) and $\bar{Z}$ runs vertically (top to bottom, $d$ Pauli-$Z$ operators), as shown in Fig.~\ref{fig:surface_code}.
Because any logical operator must span the entire lattice, at least $d$ single-qubit errors are needed to cause an undetectable logical failure.

\begin{figure}[ht]
\centering
\includegraphics[width=\columnwidth]{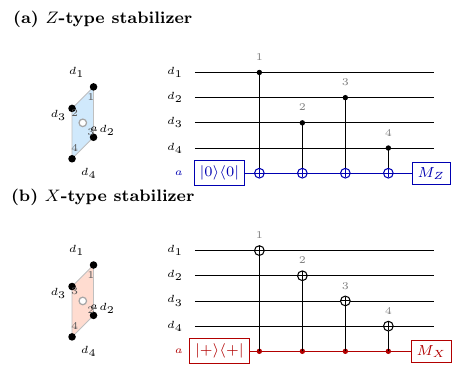}
\caption{Syndrome extraction circuits for bulk (weight-four) stabilizers.
Left: a single plaquette with four data qubits ($d_1$--$d_4$, filled) and one ancilla ($a$, open); numbers indicate the CNOT order.
Right: the corresponding circuit.
(a)~$Z$-type: ancilla prepared in $\ket{0}$, data qubits control CNOTs onto the ancilla, measured in $Z$ basis ($M_Z$).
(b)~$X$-type: ancilla prepared in $\ket{+}$, ancilla controls CNOTs onto data qubits, measured in $X$ basis ($M_X$).
Boundary stabilizers use the same circuit with only two of the four CNOT slots active.}
\label{fig:cnot_schedule}
\end{figure}

Each stabilizer acts on multiple data qubits and cannot be measured directly. Instead, an ancilla qubit is used to extract the eigenvalue of each stabilizer.
This process is called syndrome extraction, illustrated in Figure~\ref{fig:cnot_schedule} for both stabilizer types.
For $Z$-type stabilizers, the ancilla is prepared in $\ket{0}$, CNOT gates are applied with each data qubit as the control and the ancilla as the target, and the ancilla is measured in the $Z$ basis ($M_Z$).
For $X$-type stabilizers, the ancilla is prepared in $\ket{+}$, the ancilla controls CNOTs targeting each data qubit, and the ancilla is measured in the $X$ basis ($M_X$).
The outcome $s_i \in \{0,1\}$ gives the eigenvalue of stabilizer $S_i$ with $s_i = 0$ means the $+1$ eigenspace (no error detected), and $s_i = 1$ means the $-1$ eigenspace (an error anticommuting with $S_i$).

For a bulk stabilizer, the ancilla interacts with all four neighboring data qubits through four CNOT gates.
Boundary stabilizers involve only two data qubits, so two of the four CNOT slots are idle.
The gate ordering (the \emph{CNOT schedule}) must avoid having a data qubit in two gates at once.
It also matters for error propagation since a CNOT can spread a single fault into a correlated two-qubit error, and a poorly chosen schedule can let such errors mimic low-weight logical failures, reducing the effective code distance~\cite{tomita2014low}.
Figure~\ref{fig:cnot_schedule} shows a standard schedule~\cite{tomita2014low} that avoids this problem.
Each ancilla cycles through its four neighbors in a fixed order, so that both $X$- and $Z$-type stabilizers are extracted in parallel.
Together with the initial ancilla reset and the final measurement, this gives six time steps per round.

The outcomes of one extraction round for all stabilizers form the syndrome
\begin{equation}
\text{Syn}^{(k)} = \bigl(s^{(k)}_{1},\, s^{(k)}_{2},\, \ldots,\, s^{(k)}_{d^2-1}\bigr),
\end{equation}
where $s_i^{(k)}$ is the measurement result of the $i$-th stabilizer in round $k$.
If only data-qubit errors were present, a single syndrome would be enough to locate all errors.
In practice, however, the ancilla qubits and CNOT gates are themselves noisy.
For example, if an ancilla measurement flips, the syndrome bit $s_i^{(k)}$ changes even without any data error.
A single round cannot tell whether $s_i^{(k)} = 1$ was caused by a data error or a measurement error.
To resolve this, fault-tolerant protocols repeat the extraction for $d_m$ rounds (typically $d_m = d$)~\cite{dennis2002topological}, building up a full syndrome record
\begin{equation}
\text{Syn} = \bigl(\text{Syn}^{(1)},\, \text{Syn}^{(2)},\, \ldots,\, \text{Syn}^{(d_m)}\bigr)
\end{equation}
before calling the decoder.
We call the complete sequence of $d_m$ extraction rounds plus one decoder invocation a single QEC cycle.
The key idea is that a real data error persists across rounds and leaves a consistent pattern, while a measurement error typically affects only one round and can be spotted by comparing consecutive outcomes.

\subsection{Space--time syndrome volume}
\label{sec:spacetime_volume}

The repeated extraction naturally gives rise to a three-dimensional data structure that is central to the decoding problem under realistic noise.
As first recognized by Dennis~\emph{et al.}~\cite{dennis2002topological}, arranging the $d_m$ syndrome rounds along a temporal axis $t$, with the two spatial dimensions of the stabilizer lattice forming the plane at each time slice, one obtains the space--time syndrome volume.
For a distance-$d$ code measured over $d_m = d$ rounds, the $d^2-1$ stabilizers each produce $d$ syndrome bits, yielding $d(d^2-1) \approx d^3$ values in total.
These are arranged into a 3D array of shape $D \times D \times T$ under the convention of Ref.~\cite{chamberland2026fast}, with $D = d$ and $T= d_m$.

Rather than decoding the raw syndrome values directly, it is advantageous to work with detection events (also called syndrome differences), defined as~\cite{gidney2021stim,fowler2012surface}
\begin{equation}
d_{i,k} = s_{i}^{(k)} \oplus s_{i}^{(k-1)},
\label{eq:detector_event}
\end{equation}
where $\oplus$ denotes addition modulo two and $s_{i}^{(0)}$ is set to the noiseless expected value.
In the Stim formalism~\cite{gidney2021stim}, a detector is a specified set of measurement records whose XOR is deterministic in the noiseless case. A detection event fires whenever this XOR deviates from its expected value.
The key advantage of working with detection events is that errors with unbounded non-local effects in the raw syndrome become bounded local effects in the detection-event representation~\cite{gidney2021stim}.
In the absence of faults, all detection events are trivial ($d_{i,k} = 0$). A non-trivial event ($d_{i,k} = 1$) signals that at least one fault has occurred between rounds $k{-}1$ and $k$.
The entries of the space--time volume are precisely these detection events.

The space--time structure of the detection events is closely related to the physics of a random-plaquette gauge model~\cite{dennis2002topological,wang2003confinement}. Dennis~\emph{et al.}~\cite{dennis2002topological} showed that error recovery under noisy measurement is equivalent to finding a minimum-weight 1-chain in a 3D cell complex whose boundary matches the observed syndrome, with the error threshold corresponding to an order--disorder phase transition.
Figure~\ref{fig:spacetime_volume} illustrates this 3D structure for a $d=3$ code.
Each stabilizer position at each measurement round corresponds to a detector, physical faults at different stages of the syndrome extraction circuit produce non-trivial detection events with distinctive spatial and temporal signatures.
Following the taxonomy of Refs.~\cite{fowler2012surface,google2023suppressing}, these fall into three classes:
\begin{figure}[ht]
\centering
\includegraphics[width=\columnwidth]{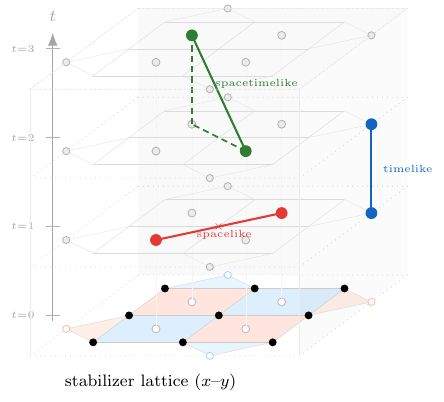}
\caption{The space--time syndrome volume for the rotated surface code ($d=3$ example).
\textbf{Base plane}: the 2D stabilizer lattice with $3\times 3$ data qubits (black dots). $Z$-type (blue) and $X$-type (orange) plaquettes alternate in a checkerboard pattern, with weight-2 boundary stabilizers (triangles) at the edges, open circles mark ancilla (detector) positions.
\textbf{Vertical axis}: measurement rounds~$t$.
At each round, every stabilizer yields a detection event $d_{i,k}$ (gray dots, $d^2{-}1 = 8$ per slice).
Three classes of faults produce distinct patterns of non-trivial events~\cite{fowler2012surface,google2023suppressing}:
\textbf{spacelike} faults (data-qubit errors, red) trigger two spatially adjacent detectors in the same time slice, \textbf{timelike} faults (measurement errors, blue) trigger the same detector in two consecutive rounds,
\textbf{spacetimelike} faults (green solid) arise from errors during the multi-step extraction circuit (most commonly entangling gates) and produce events displaced in both space and time, dashed lines show the timelike and spacelike decomposition.
A notable subclass are hook errors, where the fault propagates through subsequent gates to create a weight-2 data-qubit error.
Notice that the open circle in each time slice denotes the detector rather than the syndrome itself.}
\label{fig:spacetime_volume}
\end{figure}

\begin{itemize}
    \item \textbf{Spacelike faults} (data-qubit Pauli errors):
    A single-qubit Pauli error on a data qubit anticommutes with the surrounding stabilizers.
    Because the error persists until corrected, the syndrome-differencing operation produces a pair of non-trivial detection events at spatially adjacent detector positions within a single time slice~\cite{fowler2012surface}.
    In the matching graph, such faults correspond to spacelike edges connecting two detectors at the same temporal coordinate.

    \item \textbf{Timelike faults} (measurement and reset errors):
    A classical bit-flip during ancilla measurement or state preparation corrupts a single stabilizer outcome in round $k$ without disturbing any data qubit~\cite{google2023suppressing}.
    The syndrome-differencing operation produces a pair of detection events at the same spatial location in rounds $k$ and $k{+}1$, connected by a timelike edge in the matching graph~\cite{fowler2012surface}.
    As Terhal~\cite{terhal2015quantum} notes, timelike edges serve only to account for measurement noise and carry no physical correction.

    \item \textbf{Spacetimelike faults}:
    A fault during the multi-step syndrome extraction circuit can produce a pair of detection events displaced from each other in both space and time, with one event appearing on a different stabilizer and in a different round than the other~\cite{fowler2012surface,wang2011surface,google2023suppressing}.
    In the matching graph these appear as diagonal edges connecting detectors at different spatial and temporal positions.
    The most common source is errors associated with entangling gates. A noisy CNOT gate can induce errors on more than one qubit, including both the data qubit and the ancilla, producing detection events at different spatial and temporal coordinates.
    The direction and length of the resulting edge depend on the specific CNOT time step at which the fault occurs (see Fig.~\ref{fig:cnot_schedule}).
    Spacetimelike edges are unique to the circuit-level noise model and absent in code-capacity or phenomenological models.
    A particularly important subclass are hook errors: when a CNOT propagates fault through subsequent gates, it can create a weight-2 data-qubit error aligned with a logical operator, effectively reducing the code distance~\cite{tomita2014low,fowler2012surface}.
\end{itemize}

\noindent
This three-dimensional structure, absent in the idealized code-capacity and phenomenological noise models, is the fundamental reason that decoding the surface code under circuit-level noise is substantially harder~\cite{dennis2002topological,terhal2015quantum}.
The decoder must construct a 3D matching graph $G = (V, E)$ where each detector is a vertex and each fault mechanism contributes an edge with weight $w_e = \ln[(1-p_e)/p_e]$, then find a minimum-weight perfect matching over all flagged vertices~\cite{fowler2012surface,higgott2025sparse}.
Crucially, the graph contains spacelike, timelike, and spacetimelike edges simultaneously, and the decoder must correctly pair detection events that can be separated in space, in time, or in both dimensions.
For a distance-$d$ code, the number of non-trivial detection events scales as $s \propto p\,d^3$ (where $p$ is the physical error rate), making the computational cost of exact MWPM decoding grow rapidly with code size~\cite{higgott2025sparse}.

\subsection{Detector error models and the decoding graph}
\label{sec:dem}

The detector error model (DEM)~\cite{gidney2021stim,derks2025designing} formalizes the mapping from physical faults to detection events.
It describes the noisy circuit purely in terms of observable consequences, that is, detection events and logical frame changes, rather than the underlying Pauli errors.
Each independent fault mechanism~$f$ is characterized by three quantities~\cite{gidney2021stim}:
(i)~its probability $p(f)$,
(ii)~its detection signature $\mathcal{D}(f)$,  which is defined as the set of detectors it flips, and
(iii)~its frame change $\mathcal{L}(f)$, which is defined as the set of logical observables it flips.

Stim~\cite{gidney2021stim} constructs the DEM automatically from any stabilizer circuit with a noise model.
The procedure has two steps.
First, a noiseless stabilizer-tableau simulation produces the deterministic measurement outcomes that define the expected detector parities, this fault free outputs acts as a reference sample.
Second, for each fault location, Stim propagates the corresponding Pauli error forward through the remaining Clifford gates via Pauli frame tracking~\cite{gidney2021stim}, recording which detectors flip ($\mathcal{D}(f)$) and whether any logical observable is affected ($\mathcal{L}(f)$).
Each fault location is analyzed independently, and the propagation cost is linear in circuit depth.
The total number of DEM entries therefore scales as $\mathcal{O}(|\text{circuit}|)$, not exponentially with the number of qubits.
The DEM formalism assumes Pauli stochastic noise, in which coherent errors are approximated as probabilistic mixtures of Pauli errors~\cite{gidney2021stim}.
This is standard for matching-based decoders~\cite{dennis2002topological,higgott2025sparse}.

The DEM directly defines the weighted matching graph (also called the decoding graph) $G = (V, E)$~\cite{dennis2002topological,higgott2025sparse}, in which 
each detector $D_i$ is treated as a vertex $v_i \in V$.
A distinguished boundary vertex $v_\partial$ is also introduced.
Faults with $|\mathcal{D}(f)|=2$ produce edges between the two corresponding detector vertices.
Faults with $|\mathcal{D}(f)|=1$ produce boundary edges connecting the single detector to~$v_\partial$, these represent errors near the code boundary where one stabilizer endpoint is absent~\cite{higgott2025sparse}.
Faults triggering three or more detectors form hyperedges, which no ordinary graph can represent directly.
Stim decomposes each into a pair of graph edges~\cite{gidney2021stim}.
Every edge carries a log-likelihood-ratio weight
\begin{equation}
    w(f) \;=\; \ln\!\frac{1 - p(f)}{p(f)}\,,
    \label{eq:edge_weight}
\end{equation}
so that minimizing total matched weight is equivalent to maximum-likelihood decoding under independent faults~\cite{dennis2002topological}.
When multiple faults share the same detection signature and frame change, they merge into a single edge.
The effective probability follows the XOR combination rule $p_{\text{eff}} = p_1 + p_2 - 2p_1 p_2$, i.e., the probability that an odd number of the contributing faults occur.

The edges of the matching graph inherit the space--time structure of the syndrome volume (Sec.~\ref{sec:spacetime_volume}).
Spacelike edges connect detectors within the same time slice (data-qubit errors).
Timelike edges connect the same detector across consecutive rounds (measurement errors).
Spacetimelike edges are displaced in both space and time (faults during the syndrome extraction circuit, primarily entangling gates).
The matching graph and the 3D syndrome tensor are therefore two complementary views of the same detection event structure. The graph suits combinatorial matching algorithms, while the tensor suits convolutional neural network processing.
The DEM captures the fault-mechanism structure that Stim exposes to PyMatching, while the same underlying noise model sampled at the circuit level produces the training tensors for the neural pre-decoder (Fig.~\ref{fig:pipeline}).

Given the set of flagged detection events, MWPM finds a minimum-weight perfect matching~$M$ that pairs all flagged vertices.
The boundary vertex $v_\partial$ absorbs any unpaired defect.
The decoder's logical correction is $\hat{L} = \bigoplus_{e \in M} \mathcal{L}(e)$.
The decoder does not need to identify the exact physical error, it only needs the correct logical equivalence class.
Two error patterns that differ by a stabilizer element have identical effects on the logical state.
Decoding therefore succeeds whenever the inferred correction is homologically equivalent to the actual error, even if the matched edges differ from the true fault pattern~\cite{dennis2002topological,fowler2012surface}.

\subsection{Pre-decoding and baseline architecture}
\label{sec:predecoder_role}

The cost of MWPM decoding grows with the number of flagged detectors.
Even with the efficient sparse-blossom algorithm of PyMatching~\cite{higgott2025sparse}, global decoding remains the computational bottleneck near the error threshold.
Most physical errors produce detection events that are locally clustered (Fig.~\ref{fig:spacetime_volume}).
A pre-decoder~\cite{chamberland2026fast,meinerz2022scalable} exploits this locality by applying a CNN to local neighborhoods of the 3D tensor, resolving the majority of errors without global information.
The remaining unresolved events form the residual syndrome, which is passed to MWPM with far fewer flagged vertices.

Let $s$ denote the syndrome density, i.e., the fraction of non-trivial detection events.
Following Ref.~\cite{chamberland2026fast}, the per-cycle decoding time for standalone MWPM and for the hybrid pipeline are
\begin{align}
T_{\text{std}}  &= T_s + T_l + T_{\text{dec}}(s)\,,
\label{eq:standalone_time}\\
T_{\text{hyb}}  &= T_s + T_{l,1} + T_{\text{pre}}
                  + T_{l,2} + T_{\text{dec}}(s')\,,
\label{eq:hybrid_time}
\end{align}
where $T_s$ is the stabilizer measurement time, $T_l$ is the communication latency for standalone decoding, $T_{l,1}$ and $T_{l,2}$ are the latencies before and after the pre-decoder in the hybrid path, $T_{\text{pre}}$ is the pre-decoder inference time (independent of~$s$), and $T_{\text{dec}}(\cdot)$ is the MWPM decoding time at syndrome density $s' \ll s$.
A net speedup is achieved whenever $T_{\text{hyb}} < T_{\text{std}}$.
Because the standalone MWPM runtime scales as $\text{poly}(|\text{Syn}|) \sim \text{poly}(s \cdot d^3)$~\cite{chamberland2026fast}, the pre-decoder's density reduction $s \to s' \ll s$ translates into a proportionally larger absolute time saving at greater code distance, where the baseline is highest.

\begin{figure*}[!htbp]
\centering
\includegraphics[width=1.8\columnwidth]{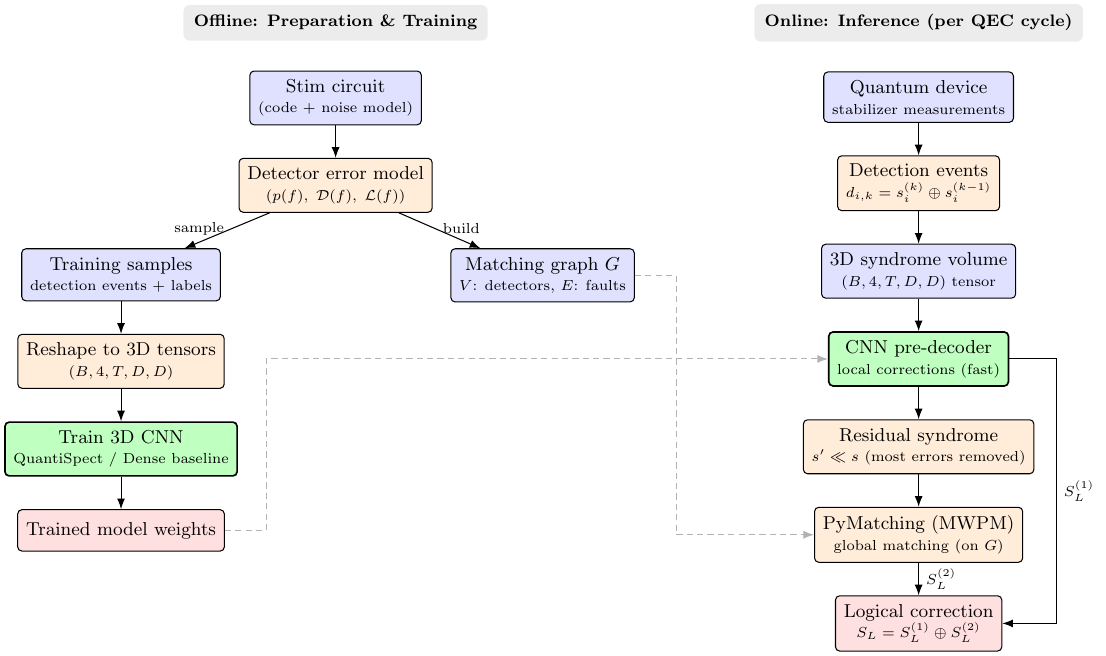}
\caption{End-to-end hybrid decoding pipeline.
\textbf{Left (offline)}: Stim constructs the detector error model (DEM) from the annotated stabilizer circuit, the DEM is used both to generate training data (detection event tensors and correction labels) and to build the matching graph $G=(V,E)$.  The 3D CNN pre-decoder is trained on the reshaped tensors.
\textbf{Right (online)}: Each QEC cycle produces a 3D syndrome volume from the stabilizer measurements.  The trained CNN outputs a sparse residual syndrome ($s' \ll s$) and its own logical prediction $S_L^{(1)}$.  PyMatching performs minimum-weight perfect matching on the pre-built graph $G$ using the residual, producing $S_L^{(2)}$.  The final logical correction is $S_L = S_L^{(1)} \oplus S_L^{(2)}$.
Dashed arrows indicate offline-to-online data flow (trained weights, pre-built graph).}
\label{fig:pipeline}
\end{figure*}

Figure~\ref{fig:pipeline} illustrates the complete hybrid pipeline.
In the offline phase, the DEM is derived from the Stim circuit.
It is used to generate CNN training data and to build the matching graph for the global decoder.
In the online phase, each QEC cycle's detection events are assembled into a 3D syndrome volume and fed to the trained CNN.
The residual syndrome is then decoded by MWPM on the matching graph.
The CNN operates on regular tensors while MWPM operates on the combinatorial graph. 

\paragraph*{Offline phase.}
Starting from a Stim circuit, the DEM is derived and used for two purposes:
(i)~generating training data (detection event tensors of shape $(B, 4, T, D, D)$, where $B$ is the batch size, $T$ the number of measurement rounds, and $D$ the code distance, paired with correction labels~\cite{chamberland2026fast}); and
(ii)~constructing the matching graph $G=(V,E)$ with weights from Eq.~\eqref{eq:edge_weight}, which is serialized for reuse at inference time.
The CNN pre-decoder is trained on these tensor pairs.

\paragraph*{Online phase.}
At each QEC cycle, the stabilizer outcomes are converted into detection events and assembled into a 3D syndrome volume.
The CNN produces local data-qubit and syndrome corrections in a single forward pass.
These predictions are combined with the original syndrome via mod-2 arithmetic to form the residual syndrome ($s' \ll s$), accounting for the induced stabilizer changes and the cumulative effect of syndrome corrections across time steps.
PyMatching~\cite{higgott2025sparse} then performs MWPM on the pre-built graph $G$ with only the residual flagged vertices.
The final logical correction combines both stages: $S_L = S_L^{(1)} \oplus S_L^{(2)}$.
The \textsc{QuantiSpect} architecture proposed in this work is a drop-in replacement for the CNN component, leaving all other pipeline elements unchanged.

Chamberland~\etal~\cite{chamberland2026fast} proposed a fully convolutional 3D neural network as the pre-decoder.
The network stacks dense Conv3D layers in sequence.
Five model variants (Models~1--5) trade off expressiveness against inference speed.
A representative configuration (Model~4, referred to as \textsc{Accurate}) has six Conv3D layers with $3{\times}3{\times}3$ kernels, each followed by Dropout3D and GeLU except the last.
The channel dimensions are $4{\to}128$ (input), $128{\to}128$ (four hidden layers), and $128{\to}4$ (output).
This gives a receptive field of $R = 1 + 6 \times (3{-}1) = 13$ and approximately 1.80\,M parameters.
Model~5 widens to $256$ filters per layer (7.13\,M parameters), while Model~1 (\textsc{Fast}) uses four layers (0.91\,M parameters).

This architecture treats the syndrome volume as a generic 3D signal, applying dense $3 \times 3 \times 3$ convolutions at every layer.
The design has two limitations relevant to scalable deployment.
First, spacelike, timelike, and spacetimelike errors have distinct physical origins and largely separate spatial and temporal signatures.
Uniform $3{\times}3{\times}3$ kernels do not exploit this separability.
The resulting models achieve high decoding accuracy but carry more parameters than necessary, leaving room to reduce training and inference cost.
Second, the plain sequential topology lacks residual connections and channel attention, which makes it harder to compress the network width without losing accuracy.

Beyond the network itself, Ref.~\cite{chamberland2026fast} introduced several data-processing techniques that are essential to pre-decoder performance.
The input encoding maps syndrome differences and stabilizer presence masks onto a $D \times D$ grid, producing a four-channel tensor of shape $(B, 4, T, D, D)$.
Homological equivalence protocols canonicalize training labels into unique representatives for both spacelike and timelike errors (Algorithm~3 therein handles the timelike case).
A fault-deferral scheme (Algorithm~2 therein) prevents artificial timelike detection events.
Decomposition rules for Y-error faults (Table~I therein) correctly split mixed spacelike--timelike signatures.
These components are orthogonal to the network architecture.
We retain all of them unchanged and focus our contribution on the neural network design.

\section{Method}
\label{sec:method}

In this section, we describe the \textsc{QuantiSpect} pre-decoder architecture and the associated training and inference methodology.
Surface code errors have partially separable spatial and temporal patterns.
The core idea is to encode this known property directly into the network architecture, so the network does not have to learn it from data the way a generic dense network would.
This yields a more parameter-efficient architecture while preserving the decoding accuracy and receptive field of the dense baseline.

\subsection{Input and output data structure}
\label{sec:io_structure}

Following Ref.~\cite{chamberland2026fast}, the input tensor $\mathbf{X}$ has shape $(B, 4, T, D, D)$, where $B$ is the batch size, $T = d_m$ is the number of syndrome measurement rounds, and $D = d$ is the code distance.
The four input channels encode both syndrome signals and structural priors:
\begin{enumerate}
    \item \textbf{Channel 1} ($x_{\text{syn}}$): $X$-type detector events mapped onto the $D \times D$ grid.
    Each weight-four (weight-two) $X$ stabilizer is placed at the top-left (top) data qubit in its support~\cite{chamberland2026fast}.
    \item \textbf{Channel 2} ($z_{\text{syn}}$): $Z$-type detector events mapped onto the same grid.
    Each weight-four (weight-two) $Z$ stabilizer is placed at the top-right (right) data qubit in its support.
    \item \textbf{Channel 3} ($x_{\text{pres}}$): $X$-stabilizer presence mask.
    Each grid cell holds 1.0 for a weight-four bulk stabilizer, 0.5 for a weight-two boundary stabilizer, or 0.0 if no stabilizer is present.
    \item \textbf{Channel 4} ($z_{\text{pres}}$): $Z$-stabilizer presence mask, with the same weighting convention.
\end{enumerate}
The first two channels carry the dynamic syndrome signal.
The latter two provide static geometric priors about the lattice, so the network does not need to learn boundary rules from data alone.
In a $Z$-basis ($X$-basis) memory experiment, $x_\text{pres}$ and $x_\text{syn}$ ($z_\text{pres}$ and $z_\text{syn}$) are set to zero at both $k=1$ and $k=d_m$, removing the irrelevant stabilizer type at the boundary rounds.

The pre-decoder output tensor $\hat{Y}$ has the same shape $(B, 4, T, D, D)$, stored as raw logits before sigmoid activation.
The input tensor holds what the decoder observes, the detection events and the stabilizer geometry.
The output tensor holds the underlying errors that produced them, both on the data qubits and in the measurements.
Its four channels encode two types of corrections:
\begin{enumerate}
    \item \textbf{Channels 1--2} (\emph{spacelike}): predicted logits for $Z$-type and $X$-type Pauli errors on data qubits.
    After thresholding, these give binary estimates of data-qubit errors between consecutive measurement rounds.
    \item \textbf{Channels 3--4} (\emph{timelike}): predicted logits for measurement bit-flip errors on $X$-type and $Z$-type syndrome qubits, respectively.
    These capture faulty ancilla readouts that corrupt a syndrome bit without any data-qubit error.
\end{enumerate}
Both correction types are needed for accurate residual syndrome construction.
Spacelike corrections identify data-qubit errors and their induced syndrome signatures.
Timelike corrections account for measurement errors that would otherwise appear as spurious detection events in the residual.
Together they let the downstream global decoder operate on a clean, sparse residual syndrome.

\subsection{Loss function}
\label{sec:loss}

The pre-decoder is trained with a binary cross-entropy (BCE) loss~\cite{chamberland2026fast}, applied independently to every output voxel across all four channels.
For a single sample, let $Y \in \{0,1\}^{4 \times T \times D \times D}$ be the ground-truth labels and $\hat{Y} \in [0,1]^{4 \times T \times D \times D}$ the sigmoid-activated model outputs ($T = d_m$).
The loss function is
\begin{equation}
\mathcal{L}_{\text{BCE}}(Y, \hat{Y}) = \sum_{c=1}^{4} \sum_{k=1}^{T} \sum_{\alpha=1}^{D} \sum_{\beta=1}^{D} \ell(Y_{c,k,\alpha,\beta},\; \hat{Y}_{c,k,\alpha,\beta}),
\label{eq:bce_loss}
\end{equation}
where
\begin{equation}
\ell(y, \hat{y}) = -y \log \hat{y} - (1-y)\log(1-\hat{y}),
\end{equation}
giving $4TD^2$ terms per sample.
In practice, the network outputs raw logits, and the sigmoid is folded into the loss computation for numerical stability.

The ground-truth labels $Y$ follow the preprocessing of Ref.~\cite{chamberland2026fast}.
Spacelike corrections (channels 1--2) are the error differences between consecutive rounds after spacelike and timelike homological equivalence.
Timelike corrections (channels 3--4) are obtained by propagating the output data-qubit errors through an additional noiseless stabilizer round (Algorithm~1 in Ref.~\cite{chamberland2026fast}) and XOR-ing the resulting syndromes $s_2$ with the original syndromes $s_1$, isolating the purely timelike failure component.
This construction ensures the network learns physically meaningful local corrections rather than artifacts of the circuit structure.

\subsection{QuantiSpect architecture}
\label{sec:architecture}

\begin{figure*}[t]
    \centering
    \includegraphics[width=\textwidth]{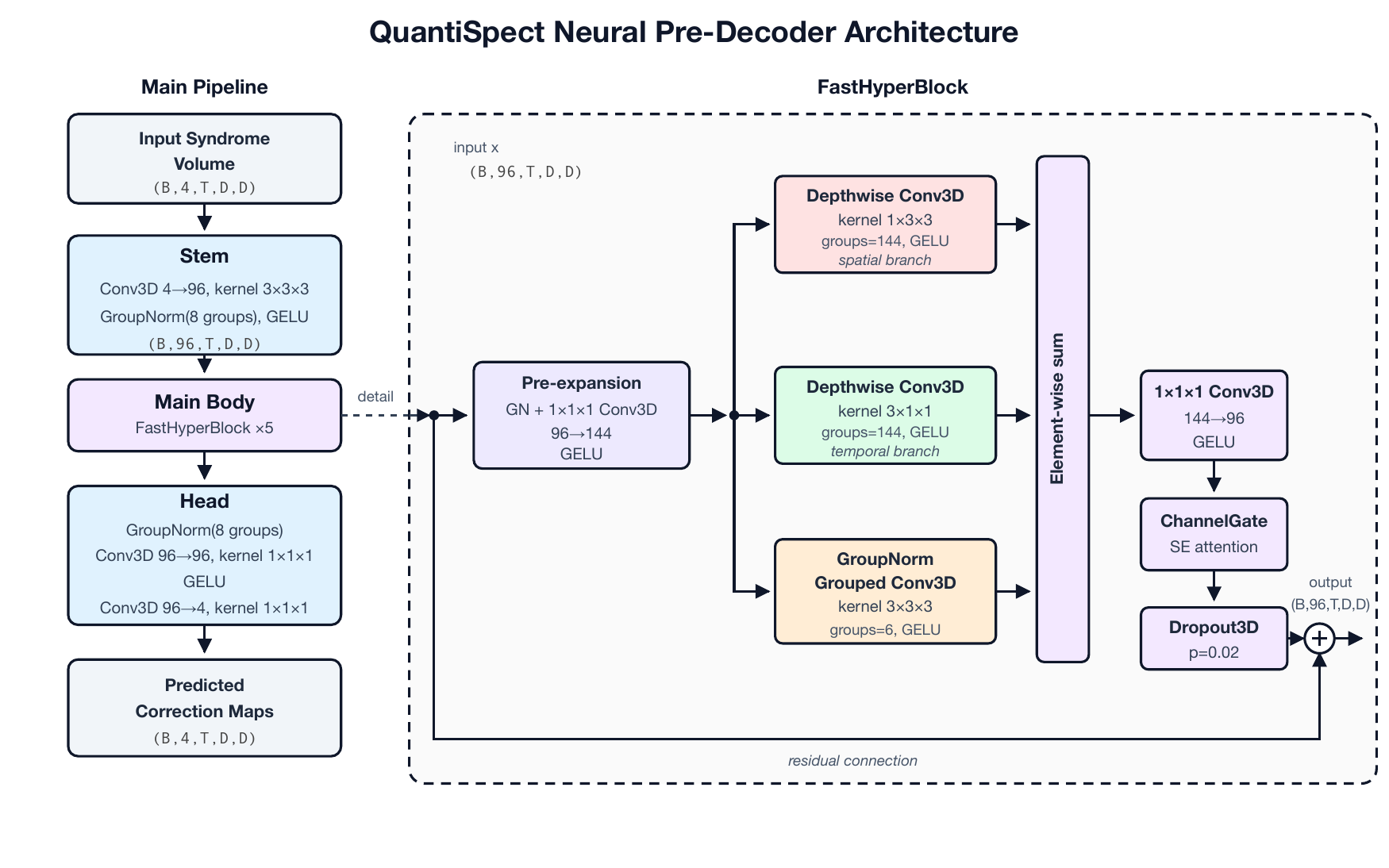}
    \caption{Architecture of the \textsc{QuantiSpect} neural pre-decoder. The input syndrome volume of shape $(B, 4, T, D, D)$ passes through a stem (Conv3D + GroupNorm + GELU), a main body of five identical FastHyperBlock residual blocks, and a head (GroupNorm + Conv3D + GELU + Conv3D) to produce the four-channel output correction maps. Each FastHyperBlock decomposes the 3D convolution into three parallel branches: \textit{spatial} (depthwise $1 \times 3 \times 3$) for local planar patterns, \textit{temporal} (depthwise $3 \times 1 \times 1$) for cross-round evolution, and \textit{mixed} (grouped $3 \times 3 \times 3$) for spatio-temporal coupling. Branches are summed, projected via a $1 \times 1 \times 1$ fusion convolution, refined by squeeze-and-excitation channel gating, and added to the input via a residual connection.}
    \label{fig:architecture}
\end{figure*}

The \textsc{QuantiSpect} architecture consists of three stages.
A stem performs initial feature extraction, a main body of five identical residual blocks performs deep feature processing, and a head remaps channels to the output space.
The complete architecture is illustrated in Fig.~\ref{fig:architecture}, and the default hyperparameters are listed in Table~\ref{tab:hyperparams}.

\begin{table}[t]
\centering
\caption{Default architectural hyperparameters for the \textsc{QuantiSpect} pre-decoder.}
\label{tab:hyperparams}
\begin{tabular}{lcc}
\toprule
\textbf{Parameter} & \textbf{Symbol} & \textbf{Value} \\
\midrule
Input / output channels & & 4 / 4 \\
Hidden dimension & $C$ & 96 \\
Mid dimension & $C_{\text{mid}}$ & 144 \\
Mixing groups & $G$ & 6 \\
Number of blocks & $N$ & 5 \\
Stem kernel size & & $3 \times 3 \times 3$ \\
Gate reduction ratio & $r$ & 4 \\
Dropout probability & $p_{\text{drop}}$ & 0.02 \\
\midrule
Total parameters & & $\sim0.663$\,M \\
Receptive field & $R$ & 13 \\
\bottomrule
\end{tabular}
\end{table}

\subsubsection{Stem}
\label{sec:stem}

The stem maps the four-channel input to a $C$-dimensional hidden space:
\begin{equation}
\text{Stem}: \text{Conv3D}(4 \to C,\ k=3) \to \text{GN}(C) \to \text{GELU},
\end{equation}
where $\text{GN}(C)$ is GroupNorm with the largest group count $g \leq 8$ that divides $C$.
For $C = 96$ this gives $g = 8$ groups of 12 channels each.
The $3 \times 3 \times 3$ kernel aggregates the local spatio-temporal neighborhood around each voxel, converting the sparse binary input into a continuous representation for the subsequent parallel-branch blocks.

\subsubsection{FastHyperBlock}
\label{sec:fasthyperblock}

The main body consists of $N = 5$ identical \texttt{FastHyperBlock} residual blocks.
Each block preserves the tensor shape $(B, C, T, D, D)$ and is the core architectural contribution of this work.

\paragraph{Pre-expansion.}
The input is expanded from $C$ to $C_{\text{mid}}$ channels via $\text{GN} \to \text{Conv3D}_{1\times1\times1}(C {\to} C_{\text{mid}}) \to \text{GELU}$.
Following the inverted bottleneck principle~\cite{sandler2018mobilenetv2}, spatial and temporal operations run in the higher-dimensional space $C_{\text{mid}}$ while the residual path stays at dimension~$C$.

\paragraph{Three parallel branches.}
The expanded features pass through three branches, each targeting a distinct error correlation in surface code syndromes:

\begin{itemize}
    \item \textbf{Spatial branch} (planar error locality):
    depthwise Conv3D with kernel $(1,3,3)$ followed by GELU.
    It acts only in the two spatial dimensions, capturing local planar syndrome patterns such as those from single data-qubit errors that trigger nearby stabilizers.
    The depthwise structure keeps each channel independent, using only $C_{\text{mid}} \times 9$ parameters.

    \item \textbf{Temporal branch} (cross-round dynamics):
    depthwise Conv3D with kernel $(3,1,1)$ followed by GELU.
    It acts only along the time axis, capturing syndrome persistence, cancellation between consecutive rounds, and measurement fault signatures, with only $C_{\text{mid}} \times 3$ parameters.

    \item \textbf{Mixed branch} (spatio-temporal coupling):
    GroupNorm $\to$ grouped Conv3D with kernel $(3,3,3)$ and $G$ groups $\to$ GELU.
    It models joint spatio-temporal correlations with controlled cross-channel mixing, and is the one that directly applies a joint spatio-temporal convolution.
    With $G=6$ and $C_{\text{mid}}=144$, each group processes 24 channels through a full $3^3$ kernel, giving a $G$-fold parameter saving over dense convolution.
\end{itemize}

\noindent
The three branches can be written as
\begin{align}
\texttt{spatial}(x)  &= \text{GELU}\!\bigl(\text{DWConv}_{1 \times 3 \times 3}(x)\bigr), \notag \\
\texttt{temporal}(x) &= \text{GELU}\!\bigl(\text{DWConv}_{3 \times 1 \times 1}(x)\bigr), \label{eq:branches} \\
\texttt{mixed}(x)    &= \text{GELU}\!\bigl(\text{GConv}_{3 \times 3 \times 3}(\text{GN}(x))\bigr), \notag
\end{align}
and their outputs are summed element-wise as $h = \texttt{spatial}(x_{\text{mid}}) + \texttt{temporal}(x_{\text{mid}}) + \texttt{mixed}(x_{\text{mid}})$.

\paragraph{Fusion, gating, and residual connection.}
A $1{\times}1{\times}1$ convolution with GELU projects $h$ back to $C$ channels ($\text{Conv3D}(C_{\text{mid}} {\to} C, k=1) \to \text{GELU}$).
A squeeze-and-excitation (SE) channel gate~\cite{hu2018squeeze} then recalibrates each channel:
\begin{equation}
\texttt{gate}(x) = \sigma\!\left(W_2\,\text{GELU}\!\left(W_1\,\text{GAP}(x)\right)\right) \odot x,
\end{equation}
where $\text{GAP}$ is global average pooling, $W_1 \in \R^{(C/r) \times C}$ and $W_2 \in \R^{C \times (C/r)}$ are learnable projections, $\sigma$ is the sigmoid, and $r=4$.
Dropout3D ($p=0.02$) and a residual connection give the block output:
\begin{equation}
\texttt{FastHyperBlock}(x) = x + \text{Dropout3D}\!\bigl(\texttt{gate}(\texttt{fuse}(h))\bigr).
\end{equation}
Each block thus learns an incremental correction to the identity mapping, suited to the pre-decoder's task of predicting sparse, local syndrome modifications.

\subsubsection{Head}
\label{sec:head}

The head maps the $C$-dimensional features back to four output channels via $\text{GN} \to \text{Conv3D}_{1\times1\times1}(C {\to} C) \to \text{GELU} \to \text{Conv3D}_{1\times1\times1}(C {\to} 4)$.
The final convolution has no activation, since the outputs are raw logits for the four correction channels (spacelike $Z$/$X$ and timelike $X$/$Z$).

\subsection{Design rationale and comparison}
\label{sec:rationale}

The three-branch design of \textsc{QuantiSpect} is motivated by the factorized structure of spatio-temporal error correlations in the surface code.

Physical errors on data qubits trigger syndrome changes on at most four neighboring stabilizers, producing compact spatial patterns on the 2D lattice.
The spatial branch ($1{\times}3{\times}3$ depthwise) captures these patterns efficiently.
Measurement errors cause syndrome persistence or cancellation across consecutive rounds.
Data-qubit errors occurring late in a measurement round may also appear only in the subsequent round's syndrome.
The temporal branch ($3{\times}1{\times}1$ depthwise) models these cross-round effects.
The residual syndrome construction in the inference procedure (Sec.~\ref{sec:inference}) provides additional motivation for the design of this branch. The residual at round~$t$ is computed from predictions at both $t$ and $t{-}1$, so the network must learn cross-round dependencies.
Some errors span both space and time simultaneously.
For example, CNOT gate failures, multi-qubit correlated faults, and burst measurement errors all produce such joint signatures.
The mixed branch ($3{\times}3{\times}3$ grouped) captures these patterns through controlled cross-channel interaction.

This factorization separates spatial and temporal mixing from cross-channel mixing.
A dense layer instead couples the two, mixing all channels across the full $3{\times}3{\times}3$ kernel at every voxel.
Avoiding that coupling is what makes \textsc{QuantiSpect} far more parameter-efficient.
The channel gate applies a sample-adaptive rescaling to each channel, emphasizing the most useful features for the current input.
GroupNorm~\cite{wu2018group} improves robustness to batch size variations.
The residual connections ensure stable gradient flow and naturally match the pre-decoder's task of learning incremental corrections.

\label{sec:receptive_field}
For a network with $l$ convolutional layers having kernel sizes $k_1, k_2, \ldots, k_l$ (all with unit stride and no dilation), the receptive field is~\cite{chamberland2026fast}
\begin{equation}
R = 1 + \sum_{i=1}^{l} (k_i - 1).
\label{eq:receptive_field}
\end{equation}
For \textsc{QuantiSpect}, the contributions come from the stem ($k=3$), $N$ blocks (each expanding the RF by $k=3$ via the parallel branches), and the head ($k=1$):
\begin{equation}
R = 1 + \underbrace{(3-1)}_{\text{stem}} + \underbrace{N \times (3-1)}_{N\text{ blocks}} + \underbrace{(1-1)}_{\text{head}} = 1 + 2 + 2N.
\end{equation}
With $N=5$ blocks, this gives $R = 13$, matching the \textsc{Accurate} baseline~\cite{chamberland2026fast}.
We call this base model \textsc{QuantiSpect}-13, with $N=5$ blocks, $R=13$, and $0.66$\,M parameters.
Because the block design is modular, the receptive field extends by simply adding blocks.
We use this property in Sec.~\ref{sec:rf_scaling} to build a deeper variant, \textsc{QuantiSpect}-21, and study how it scales.
Table~\ref{tab:comparison} summarizes how \textsc{QuantiSpect}-13 differs from the baselines.
It matches the $R=13$ receptive field of \textsc{Accurate} while using fewer parameters than any baseline.

\begin{table}[!htbp]
\centering
\caption{Architectural comparison between \textsc{QuantiSpect}-13 and the baseline models from Ref.~\cite{chamberland2026fast}.}
\label{tab:comparison}
\small
\begin{tabular}{lccc}
\toprule
 & \textbf{Fast} & \textbf{Accurate} & \textbf{QS-13} \\
\midrule
Architecture & Dense & Dense & Factor. \\
Receptive field $R$ & 9 & 13 & 13 \\
Hidden width & 128 & 128 & 96 \\
Layers / blocks & 4 lay. & 6 lay. & 5 blk. \\
Normalization & None & None & GN \\
Residual conn. & No & No & Yes \\
Channel gating & No & No & SE \\
ST decomp. & No & No & Yes \\
\bottomrule
\end{tabular}
\end{table}

\subsection{Parameter and computational cost}
\label{sec:complexity}
\label{sec:mac}

The pre-decoder's cost has two sides.
Its parameter count sets the memory footprint, while its per-voxel count of multiply-accumulate operations, the MACs, sets the inference compute.
These two nearly coincide for the stride-1 convolutions used here, because each weight contributes one MAC at every voxel.

Table~\ref{tab:param_detail} breaks the parameter count down by component for \textsc{QuantiSpect} and the two baselines.
The baselines are flat stacks of Conv3D layers, \textsc{Fast} reaching $R=9$ and \textsc{Accurate} $R=13$, whereas \textsc{QuantiSpect} is built from a stem, $N$ residual blocks, and a head.

\begin{table}[!htbp]
\centering
\caption{Parameter count breakdown by component.
\textsc{Fast} and \textsc{Accurate} are flat sequential Conv3D networks. \textsc{QuantiSpect} uses factorized residual blocks.}
\label{tab:param_detail}
\small
\setlength{\tabcolsep}{3pt}
\begin{tabular}{lrrr}
\toprule
\textbf{Component} & \textbf{Fast} & \textbf{Accurate} & \textbf{QS-13} \\
\midrule
Input / Stem & 13,952 & 13,952 & 10,656 \\
Body (per layer/blk) & 442,496 & 442,496 & 128,568 \\
Body (total) & 884,992 & 1,769,984 & 642,840 \\
Output / Head & 13,828 & 13,828 & 9,892 \\
\midrule
\textbf{Total} & $\sim$0.91\,M & $\sim$1.80\,M & $\sim$0.66\,M \\
\bottomrule
\end{tabular}
\end{table}

\textsc{Fast} and \textsc{Accurate} share the same dense building block, a $128 \to 128$ convolution with $k=3$ that costs 442496 parameters, and they differ only in how many hidden layers they stack, 2 for \textsc{Fast} and 4 for \textsc{Accurate}.
This puts their totals at 0.91\,M for $R=9$ and 1.80\,M for $R=13$.
\textsc{QuantiSpect} replaces each such dense layer with a factorized residual block of only 128568 parameters, about $3.4\times$ cheaper.
Even with 5 blocks it therefore lands at just 0.66\,M parameters, below the $R=9$ \textsc{Fast} model and  $2.71\times$ fewer than those of \textsc{Accurate} while still matching its $R=13$ receptive field.

The same efficiency carries over to computation.
A dense $\text{Conv3D}(C_{\text{in}} \to C_{\text{out}},\ k)$ layer costs $C_{\text{in}} \times C_{\text{out}} \times k^3$ MACs per voxel, whereas a depthwise convolution needs only $C \times k_t \times k_h \times k_w$ and a grouped convolution with $G$ groups only $C_{\text{in}} \times C_{\text{out}} \times k^3 / G$.
The FastHyperBlock builds its branches from these cheaper convolutions, so its per-voxel cost falls well below a dense layer's.
Table~\ref{tab:mac} gives the breakdown for one block.

\begin{table}[!htbp]
\centering
\caption{Per-voxel convolutional MAC breakdown of one FastHyperBlock.}
\label{tab:mac}
\begin{tabular}{lr}
\toprule
\textbf{Component} & \textbf{Conv MACs / voxel} \\
\midrule
Pre-expansion ($96 \to 144$, $k=1$) & 13,824 \\
Spatial DW ($144$, $k=(1,3,3)$) & 1,296 \\
Temporal DW ($144$, $k=(3,1,1)$) & 432 \\
Mixed GN+GConv ($144$, $k=3$, $g=6$) & 93,312 \\
Fusion ($144 \to 96$, $k=1$) & 13,824 \\
\midrule
\textbf{Block total} & 122,688 \\
\bottomrule
\end{tabular}
\end{table}

A single block requires $122688$ MACs per voxel, representing an approximately $3.6\times$ reduction relative to a dense layer, which requires $442368$ MACs per voxel. Combining the stem, body, and head yields the model-level computational costs reported in Table~\ref{tab:mac_model}. Despite employing five blocks, compared with four dense layers in \textsc{Accurate}, to achieve the same receptive field of $R=13$, \textsc{QuantiSpect}-13 requires approximately $2.84\times$ fewer convolutional MACs per voxel.

\begin{table}[!htbp]
\centering
\caption{Model-level per-voxel convolutional MAC breakdown.
Each dense layer costs 442,368 MACs per voxel and each FastHyperBlock 122,688 (Table~\ref{tab:mac}), the remaining rows are the input and output convolutions of each network (the stem and head in \textsc{QuantiSpect}).
\textsc{QuantiSpect}-13 uses $5$ blocks to match the $R=13$ receptive field of \textsc{Accurate}'s $4$ dense body layers, yet needs $\sim2.84\times$ fewer MACs overall.}
\label{tab:mac_model}
\begin{tabular}{lrr}
\toprule
\textbf{Component} & \textbf{Accurate} & \textbf{QS-13} \\
\midrule
Input / Stem & 13,824 & 10,368 \\
Body (per layer/blk) & 442,368 & 122,688 \\
Body (total) & 1,769,472 & 613,440 \\
Output / Head & 13,824 & 9,600 \\
\midrule
\textbf{Total} & 1,797,120 & 633,408 \\
\bottomrule
\end{tabular}
\end{table}

Because the per-voxel computational costs closely scale with the corresponding parameter counts, the model-level MAC totals exhibit the same trend as the parameter totals reported in Table~\ref{tab:param_detail}. These computational and parameter efficiencies translate directly into reduced memory consumption, lower inference latency, and decreased training cost. Table~\ref{tab:rf_nvidia} provides a comparison with all five model variants introduced in Ref.~\cite{chamberland2026fast}, which cover receptive fields ranging from $R=9$ to $R=17$ and parameter counts ranging from $0.91\,M$ to $7.13\,M$.

\subsection{Inference procedure}
\label{sec:inference}

At inference time, the pre-decoder converts its output into binary corrections.
These corrections are then post-processed to build a residual syndrome for the global decoder.
The procedure follows Ref.~\cite{chamberland2026fast} and consists of five stages.

\paragraph{Forward pass and thresholding.}
Given the input tensor $\mathbf{X} \in \mathbb{R}^{4 \times T \times D \times D}$, the network produces logits $\hat{\mathbf{Y}} = f_\theta(\mathbf{X})$ of identical shape.
A probability threshold of $0.5$ is applied to the sigmoid of each logit to obtain binary predictions.
The spacelike channels (channels~1--2) yield $\hat{c}_Z, \hat{c}_X \in \{0,1\}^{T \times D \times D}$, representing estimated data-qubit Pauli corrections at each round.
The timelike channels (channels~3--4) yield $\hat{m}_X, \hat{m}_Z \in \{0,1\}^{T \times D \times D}$, representing estimated measurement bit-flip corrections.

\paragraph{Induced syndrome via parity check.}
The spacelike corrections are mapped to the stabilizer basis using the parity-check matrices.
For each round~$k$, the flattened $Z$-correction grid $\hat{c}_Z^{(k)} \in \{0,1\}^{D^2}$ is multiplied by $H_X \in \{0,1\}^{n_s \times D^2}$ to produce the $X$-stabilizer induced syndrome.
Likewise, $\hat{c}_X^{(k)}$ is multiplied by $H_Z$:
\begin{equation}
\label{eq:induced_syndrome}
S_X^{(k)} = H_X \, \hat{c}_Z^{(k)} \!\!\pmod{2}, \quad
S_Z^{(k)} = H_Z \, \hat{c}_X^{(k)} \!\!\pmod{2},
\end{equation}
where $n_s = (D^2-1)/2$ is the number of stabilizers of each type.
The mapping is cross-type. $Z$-corrections produce $X$-stabilizer syndromes and $X$-corrections produce $Z$-stabilizer syndromes. This reflects the Pauli commutation structure.

As described in Sec.~\ref{sec:io_structure}, the network operates on a padded $D \times D$ grid, but only $n_s$ positions correspond to physical stabilizers.
We extract these valid positions from the timelike predictions $\hat{m}_X^{(k)}, \hat{m}_Z^{(k)}$ using index-select matrices $G_X, G_Z \in \{0,1\}^{n_s \times D^2}$.
Each matrix has exactly one nonzero entry per row, and maps the padded grid into the $n_s$-dimensional stabilizer ordering.

\paragraph{Residual syndrome construction.}
The residual syndrome captures what the pre-decoder failed to explain.
Let $\Delta_X^{(k)}, \Delta_Z^{(k)} \in \{0,1\}^{n_s}$ denote the original detector values at round~$k$.
Let $\hat{m}_X^{(k)}, \hat{m}_Z^{(k)}$ denote the gathered timelike predictions.
The residual for each stabilizer type is
\begin{align}
\label{eq:residual}
R^{(1)} &= \bigl(\Delta^{(1)} \oplus \hat{m}^{(1)} \oplus S^{(1)}\bigr) \bmod 2, \nonumber\\
R^{(k)} &= \bigl(\Delta^{(k)} \oplus \hat{m}^{(k)} \oplus \hat{m}^{(k-1)} \oplus S^{(k)}\bigr) \bmod 2, \quad k \geq 2,
\end{align}
where the type subscript ($X$ or $Z$) is omitted for clarity.
The $\hat{m}^{(k-1)}$ term appears because $\Delta^{(k)}$ is the difference between consecutive raw syndromes.
A measurement bit-flip at round~$k{-}1$ affects both $\Delta^{(k-1)}$ and $\Delta^{(k)}$, so the correction at $k{-}1$ must also be subtracted from round~$k$.

The resulting residuals $R_X, R_Z \in \{0,1\}^{n_s \times T}$ are rearranged to match the detector ordering used by the DEM.
First come the boundary detectors of the matching basis at round~1, then the $X$ and $Z$ stabilizers alternate at each subsequent round, and finally the terminal boundary detectors from the raw input are appended.

\paragraph{Pre-decoder logical sign.}
The pre-decoder also computes a partial logical observable flip from its spacelike corrections.
Let $\ell \in \{0,1\}^{D^2}$ encode the support of the logical operator matching the measurement basis ($\bar{X}$ for $X$-basis, $\bar{Z}$ for $Z$-basis).
The pre-decoder logical sign is
\begin{equation}
\label{eq:pre_L}
S_L^{(\mathrm{pre})} = \bigoplus_{k=1}^{T} \bigl(\ell^\top \hat{c}^{(k)}\bigr) \bmod 2,
\end{equation}
where $\hat{c}^{(k)}$ is $\hat{c}_Z^{(k)}$ for $X$-basis or $\hat{c}_X^{(k)}$ for $Z$-basis decoding.
This accumulates the parity of predicted corrections along the logical string across all rounds.

\paragraph{Global decoding and final outcome.}
The residual syndrome is passed to PyMatching~\cite{higgott2025sparse}, which runs minimum-weight perfect matching on the decoding graph (Sec.~\ref{sec:dem}) to find a correction that fits the residual.
PyMatching returns a global logical sign $S_L^{(\mathrm{global})}$ that indicates whether its correction anticommutes with the logical observable.
The final logical outcome is
\begin{equation}
\label{eq:final_logical}
S_L = S_L^{(\mathrm{pre})} \oplus S_L^{(\mathrm{global})}.
\end{equation}
A logical error is declared when $S_L$ disagrees with the true value $S_L^{(\mathrm{true})}$ recorded during simulation.
The pre-decoder and global decoder handle complementary information.
The pre-decoder removes dense, local errors through learned spatio-temporal patterns.
PyMatching resolves the sparse, long-range correlations that remain.

\subsection{Training details}
\label{sec:training_details}

Training follows the protocol of Ref.~\cite{chamberland2026fast}.
We summarize the key elements below and list all hyperparameters in Table~\ref{tab:training}.

\paragraph{Data generation and noise model.}
Training data is generated on-the-fly by a GPU-accelerated Pauli frame simulator built on Stim~\cite{gidney2021stim}.
Each sample corresponds to one QEC cycle under the circuit-level depolarizing noise model of Ref.~\cite{chamberland2026fast}, parameterized by a single physical error rate~$p$.
\begin{itemize}
    \item \emph{State preparation.} Each $\ket{0}$ ($\ket{+}$) reset is followed by an $X$ ($Z$) error with probability $2p/3$.
    \item \emph{Measurement.} Before each $Z$-basis ($X$-basis) measurement, an $X$ ($Z$) error is applied with probability $2p/3$.
    \item \emph{Two-qubit gates.} With probability $p$, a two-qubit Pauli error drawn uniformly from $\{I,X,Y,Z\}^{\otimes 2}\!\setminus\!\{I{\otimes}I\}$ (each of the 15 non-trivial terms with probability $p/15$) is applied after each CNOT.
    \item \emph{Idle locations.} Qubits idling during CNOT layers receive a single-qubit Pauli error from $\{X,Y,Z\}$ with probability $p/3$ each. Data qubits idling during the ancilla reset window receive the same error type at the higher rate $2p/3 - 2p^2/9$ each, reflecting the longer idle duration.
\end{itemize}
During training, the 25 noise parameters are upscaled to a fixed effective rate $p \approx 0.006$, near the surface code threshold.
The simulator alternates between $X$- and $Z$-basis measurements at each step to cover both stabilizer types.

\paragraph{Label construction via homological equivalence.}
The raw training labels are binary correction patterns.
Many different patterns correct the same error equivalently, since they differ only by stabilizers.
Without a consistent choice, one syndrome would map to several valid labels and send conflicting gradients to the network.
We therefore canonicalize the labels with the spacelike and timelike homological equivalence (HE) protocols of Ref.~\cite{chamberland2026fast}, following Algorithms~1--3 therein.
Spacelike HE reduces each correction pattern to a canonical representative within its homology class.
It does this by greedily lowering the correction weight stabilizer by stabilizer.
Timelike HE similarly resolves equivalent temporal correction patterns.
Both protocols are implemented in PyTorch and run on GPU.
They add negligible latency to data generation.

\paragraph{Optimization.}
We use the Lion optimizer~\cite{chen2023symbolic}.
It is memory-efficient because it keeps only a single momentum buffer per parameter and updates each weight using the sign of that momentum.
The learning rate warms up linearly over 100 steps, then drops by a factor of 0.7 at 25\%, 50\%, and 100\% of training.
Gradients are accumulated over 2 micro-batches and clipped to $\|\nabla\|_2 \leq 1.0$ before each optimizer step.
Training uses FP32 precision with TF32 tensor cores enabled.

An exponential moving average (EMA) of the model weights is maintained throughout training with $\theta_{\rm EMA} \leftarrow 10^{-4}\,\theta_{\rm EMA} + 0.9999\,\theta$.
The EMA model is used for all evaluation and inference.

\paragraph{Training window and generalization.}
The model is trained at a fixed distance $d = d_m = 13$, matching the receptive field of the architecture (Sec.~\ref{sec:receptive_field}).
Each epoch processes about $33.6$\,M samples.
The per-GPU batch size is 256 in the first epoch and 512 thereafter.
Training runs for 100 epochs on $4 \times$ A100 (40\,GB) GPUs with PyTorch DistributedDataParallel.

The architecture is fully convolutional and has no distance-dependent parameters.
The trained model therefore generalizes to arbitrary distances $d$ and round counts $T$ at inference time without retraining.
A single checkpoint serves all deployment configurations.
The model has only $0.66$\,M parameters, about $2.5$\,MB in FP32.
This small size should make it feasible to run on modest GPUs, though the actual memory footprint still depends on batch size, code distance, and intermediate activations.

\begin{table}[!htbp]
\centering
\caption{Training hyperparameters for \textsc{QuantiSpect}. Noticed that this is the configuration of $R=13$ variant, while the variant-specific changes for $R=21$ are given in Sec.~\ref{sec:rf_scaling}}
\label{tab:training}
\small
\begin{tabular}{@{}ll@{}}
\toprule
\textbf{Hyperparameter} & \textbf{Value} \\
\midrule
Training distance / rounds & $d = d_m = 13$ \\
Shots per epoch & ${\approx}33.6$\,M \\
Epochs & 100 \\
Batch size / GPU & Ep.\,1: 256; Ep.\,$\geq$2: 512 \\
Gradient accumulation & 2 steps \\
GPUs & $4 \times$ A100 (40\,GB) \\
Optimizer & Lion~\cite{chen2023symbolic} \\
 & $\beta_1=0.9,\;\beta_2=0.95,\;\text{wd}=10^{-7}$ \\
Learning rate & $2 \times 10^{-4}$ \\
LR schedule & Warmup 100 steps + step decay \\
 & $\gamma=0.7$ at 25\%/50\%/100\% \\
Loss & BCE w/ logits (sum) \\
Activation & GELU \\
Dropout & 0.02 \\
EMA update & $\theta_{\rm EMA}\!\leftarrow\!10^{-4}\,\theta_{\rm EMA}+(1{-}10^{-4})\,\theta$ \\
Precision & FP32 with TF32 enabled \\
Gradient clipping & $\|\nabla\|_2 \leq 1.0$ \\
Label protocol & HE: spacelike + weight-1 timelike \\
\bottomrule
\end{tabular}
\end{table}

\section{Numerical Results}
\label{sec:results}

In this section, we present numerical simulation results comparing the \textsc{QuantiSpect} pre-decoder with the baseline dense CNN architectures from the Ising Decoder framework~\cite{chamberland2026fast}.
We compare three models.
\textsc{Fast} and \textsc{Accurate} are Models~1 and~4 of Ref.~\cite{chamberland2026fast}, optimized for inference speed and decoding accuracy respectively.
\textsc{QuantiSpect} is the $R=13$ factorized model proposed here and described in Sections~\ref{sec:architecture}--\ref{sec:rationale}.
Their architectures and parameter counts are summarized in Table~\ref{tab:comparison}.
All experiments use identical conditions.
Circuit-level depolarizing noise is generated by Stim~\cite{gidney2021stim}, and uncorrelated PyMatching~\cite{higgott2025sparse} serves as the global decoder.
We run $N = 50000$ independent decoding shots per physical error rate for each of the $X$ and $Z$ bases.
The number of measurement rounds is $T = d$, and the code distances are $d \in \{5, 7, 9, 11, 13, 15, 17, 19, 21, 23\}$.
The physical error rate $p$ is swept over $\{0.1\%, 0.2\%, 0.3\%, \ldots, 0.8\%\}$.
Unless otherwise stated, all logical error rates reported below are the basis-averaged values $P_L^{\text{Avg}} = (P_L^{(X)} + P_L^{(Z)}) / 2$.
Statistical uncertainty is quantified using the Wilson score interval~\cite{wilson1927probable} at the 95\% confidence level.
This interval remains well behaved for the small failure counts seen at low error rates, where the usual normal approximation fails.
For each basis it has a lower bound $L$ and an upper bound $U$.
We denote these $L_X, U_X$ for the $X$ basis and $L_Z, U_Z$ for the $Z$ basis.
The interval for the basis-averaged rate is then $[(L_X + L_Z)/2,\, (U_X + U_Z)/2]$.

\subsection{Threshold and finite-size scaling}
\label{sec:threshold_curves}

\begin{figure}[tb]
\centering
\includegraphics[width=\columnwidth]{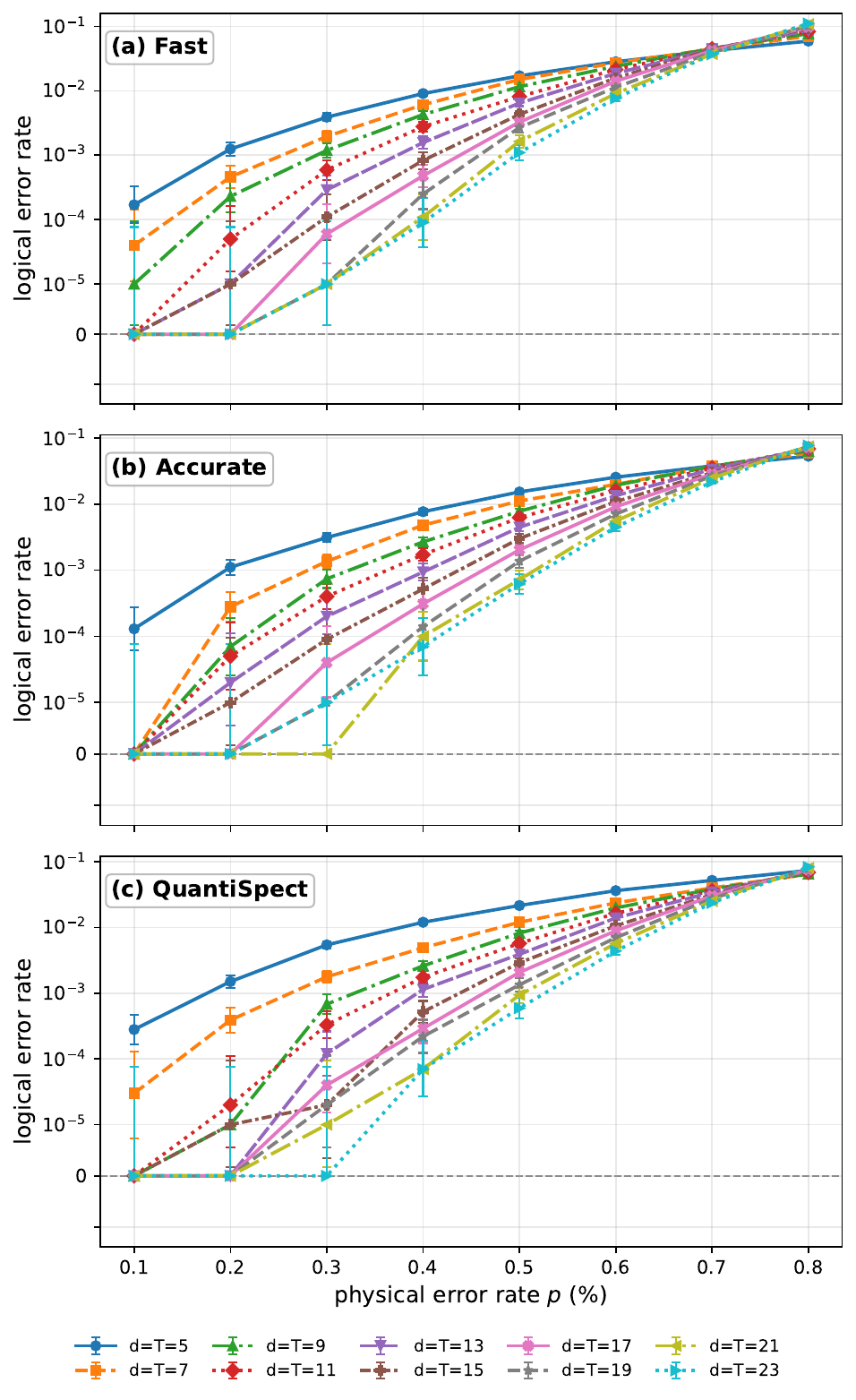}
\caption{Logical error rate vs.\ physical error rate for each pre-decoder model individually, across code distances $d \in \{5, 7, \ldots, 23\}$ with $T = d$ measurement rounds.
Panels: (a) \textsc{Fast}, (b) \textsc{Accurate}, (c) \textsc{QuantiSpect}.
All models are combined with uncorrelated PyMatching as the global decoder.
Error bars show 95\% Wilson confidence intervals ($N = 50000$ shots per basis).
The vertical axis uses a symmetric-log scale to display both positive LER values and exact-zero points (dashed baseline).}
\label{fig:threshold_individual}
\end{figure}

Figure~\ref{fig:threshold_individual} presents the logical error rate (LER) as a function of physical error rate for each of the three models individually, across all ten code distances, with 95\% Wilson confidence intervals derived from $N = 50000$ shots per basis-specific data point.

All three models exhibit clear threshold-like behavior.
Below the crossing region, increasing the code distance lowers the LER.
Above it, increasing the distance raises the LER instead.
For physical error rates $p \leq 0.2\%$, several large-distance curves reach zero observed logical failures within $50000$ shots.
This indicates strong error suppression.
The fan-out of curves is wider for \textsc{QuantiSpect} and \textsc{Accurate} than for \textsc{Fast}.
This reflects stronger per-distance error suppression in the sub-threshold regime.

To quantify the threshold, we perform a finite-size scaling (FSS) analysis using the standard ansatz~\cite{dennis2002topological}
\begin{equation}
\label{eq:fss}
\log_{10} P_L(p, d) = f\!\left((p - p_{\text{th}})\, d^{1/\nu}\right),
\end{equation}
where $p$  and $P_L$ is the physical and logical error rates respectively, $f$ is approximated by a polynomial of order~2, $p_{\text{th}}$ is the threshold error rate, and $\nu$ is the correlation-length critical exponent.
We fit this ansatz over the near-threshold window $p \in [0.3\%,\, 0.8\%]$ with $d \geq 7$.
This avoids the far-from-threshold tails, where the polynomial ansatz breaks down.
It also avoids the small-$d$ regime, where the factorized architecture behaves atypically (Sec.~\ref{sec:ler_comparison}).
Within this window, we exclude zero-LER points, since $\log_{10}(0)$ is undefined.
We then fit with Wilson-weighted least squares.
The standard deviation of $\log_{10}(P_L)$ from the Wilson 95\% confidence interval is estimated, so that points with fewer observed logical failures contribute less to the fit.
Appendix~\ref{sec:appendix_fss_scan} compares this weighting against an unweighted fit and an event-count-based robustness check, and shows that it does not change the ranking between methods.

\begin{table}[!htbp]
\centering
\caption{Circuit-level threshold estimates $p_{\text{th}}$ from Wilson-weighted finite-size scaling fits over the near-threshold window $p \in [0.3\%, 0.8\%]$ with $d \geq 7$.
Uncertainties are from the curve-fit covariance and do not include systematic fitting-window effects.
RMSE is the collapse residual in $\log_{10}$ space.}
\label{tab:fss_threshold}
\begin{tabular}{l c c c}
\toprule
\textbf{Method} & $p_{\text{th}}$ (\%) & $\nu$ & RMSE \\
\midrule
Fast        & $0.740 \pm 0.002$ & $1.29$ & $0.43$ \\
Accurate    & $0.769 \pm 0.002$ & $1.26$ & $0.33$ \\
QuantiSpect & $0.769 \pm 0.002$ & $1.25$ & $0.36$ \\
\bottomrule
\end{tabular}
\end{table}

Table~\ref{tab:fss_threshold} summarizes the FSS results.
\textsc{QuantiSpect} and \textsc{Accurate} are essentially indistinguishable at $p_{\text{th}} = 0.769\%$, and both are higher than \textsc{Fast} at $0.740\%$.
The fitted critical exponents span $\nu \approx 1.25$--$1.29$.
Appendix~\ref{sec:appendix_fss_scan} shows that an unweighted fit gives a lower $p_{\text{th}}$ for every method, by $0.03$--$0.06$ percentage points.
That fit also produces a substantially tighter FSS collapse.
The ranking between methods is the same under both weightings.

We emphasize that the quoted uncertainties reflect only the statistical precision of the curve-fit covariance.
They do not account for systematic sensitivity to fitting choices such as the polynomial order, the fitting window, and the minimum included distance.
We therefore ran a robustness scan over 80 fitting configurations per method.
The scan covers polynomial orders 2 and 3, five $p$-range windows, four minimum-distance cutoffs, and both weighted and unweighted modes.
Appendix~\ref{sec:appendix_fss_scan} gives the full parameter grid.
The resulting threshold ranges are $0.55\%$--$0.74\%$ for \textsc{Fast}, $0.67\%$--$0.78\%$ for \textsc{Accurate}, and $0.60\%$--$0.78\%$ for \textsc{QuantiSpect}.
Across the majority of configurations, \textsc{QuantiSpect} and \textsc{Accurate} remain close and both remain above \textsc{Fast}.

To check whether the extracted $p_{\text{th}}$ and $\nu$ are meaningful, we examine the FSS data collapse.
If the scaling ansatz holds, curves for different code distances should fall onto a single function $f$ when plotted against the rescaled variable $(p - p_{\text{th}})\,d^{1/\nu}$.
Appendix~\ref{sec:appendix_fss_scan} plots the collapse for both weighting schemes.
All three methods collapse reasonably well, with RMSE of $0.33$--$0.43$ for the Wilson-weighted fit and $0.14$--$0.17$ for the unweighted fit in $\log_{10}$ space.
The larger Wilson-weighted RMSE does not signal a worse fit.
The RMSE is a plain unweighted average, so it rewards the many low-$p$ points that the unweighted fit follows closely.
The Wilson-weighted fit instead puts its weight on the high-$p$ region near the threshold, where the failure counts are large and the estimates are statistically tight.
Because that region determines $p_{\text{th}}$, the Wilson-weighted collapse is the more appropriate basis for the threshold estimate.
Its higher plain RMSE reflects this weighting rather than a poorer fit.

\subsection{Comparison across code distances}
\label{sec:ler_comparison}

\begin{figure*}[t]
\centering
\includegraphics[width=\textwidth]{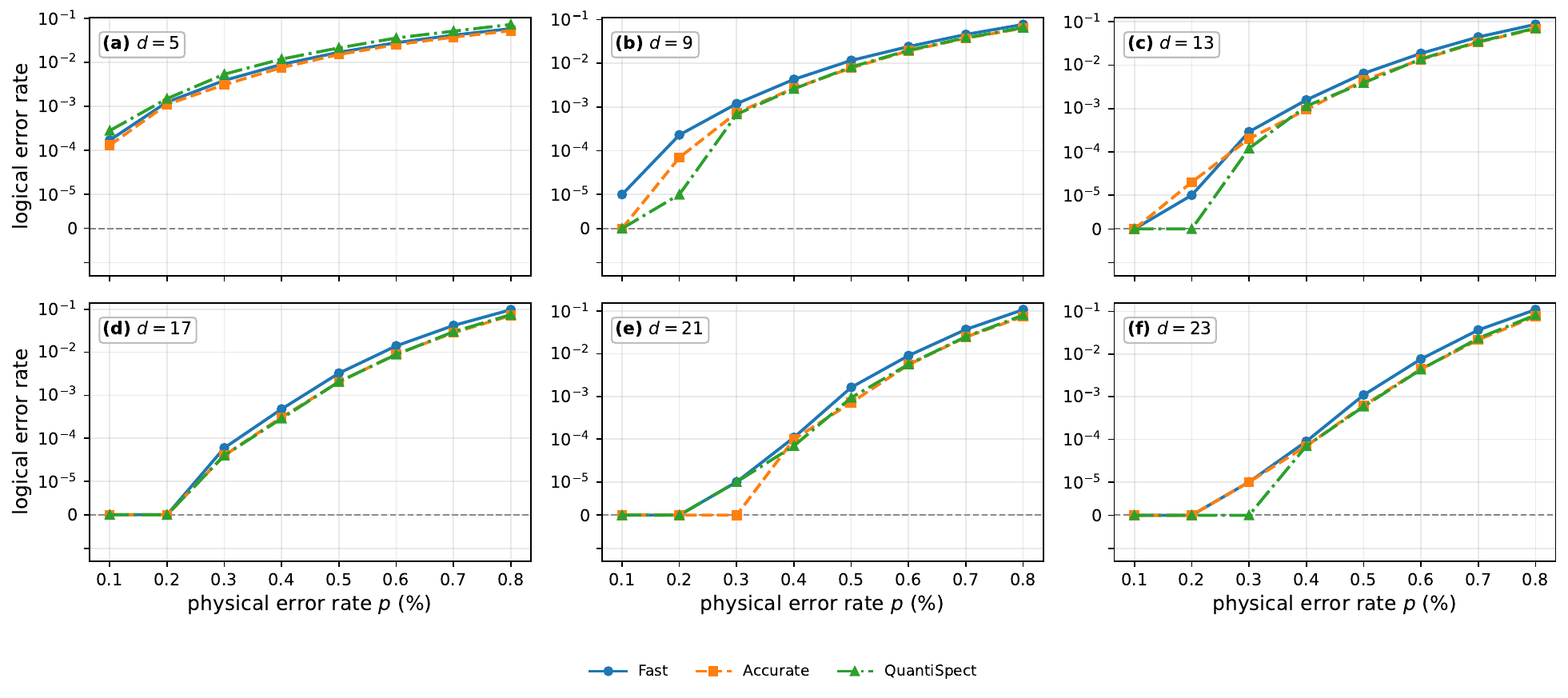}
\caption{Head-to-head comparison of logical error rate at six representative code distances, panels (a)--(f) for $d = 5, 9, 13, 17, 21, 23$ (each with $T = d$).
Three methods are compared: \textsc{Fast} (blue circles, solid), \textsc{Accurate} (orange squares, dashed), and \textsc{QuantiSpect} (green triangles, dash-dotted).
All models are evaluated under identical conditions (50000 shots per basis per point, uncorrelated PyMatching global decoder).
The vertical axis uses a symmetric-log scale, the dashed horizontal line marks $\text{LER} = 0$.}
\label{fig:compare_selected}
\end{figure*}

To compare the three models directly, we evaluated their logical error rate at six representative code distances spanning the full range ($d = 5$--$23$).
The results are shown in Figure~\ref{fig:compare_selected}.
The complete set for all ten distances appears in Figure~\ref{fig:compare_all} in the Appendix.
Table~\ref{tab:ler_comparison} lists the numerical LER values at a representative operating point $p = 0.5\%$.
It also gives the reduction factor $\rho = P_L^{\text{no\,predecoder}} / P_L^{\text{after\,predecoder}}$, which measures how much each model improves over MWPM-only decoding.

\begin{table}[!htbp]
\centering
\caption{Logical error rate (LER) after pre-decoding at $p = 0.5\%$ (Avg basis, $N = 50000$ shots per basis).
The reduction factor $\rho = P_L^{\text{no\,predec}} / P_L^{\text{after\,predec}}$ quantifies each model's error-correction effectiveness.
Each model is evaluated in an independent simulation run, so the no-predecoder baseline $P_L^{\text{no\,predec}}$ may differ slightly across methods due to finite-sample variance.
Boldface marks the best after-predecoder LER at each distance.}
\label{tab:ler_comparison}
\begin{tabular}{c c c c c c c}
\toprule
& \multicolumn{2}{c}{\textbf{Fast}} & \multicolumn{2}{c}{\textbf{Accurate}} & \multicolumn{2}{c}{\textbf{QuantiSpect}} \\
\cmidrule(lr){2-3} \cmidrule(lr){4-5} \cmidrule(lr){6-7}
$d$ & LER & $\rho$ & LER & $\rho$ & LER & $\rho$ \\
\midrule
5  & 1.71e-2 & 1.24 & \textbf{1.53e-2} & 1.46 & 2.15e-2 & 1.01 \\
7  & 1.50e-2 & 1.19 & \textbf{1.11e-2} & 1.62 & 1.20e-2 & 1.50 \\
9  & 1.16e-2 & 1.17 & \textbf{7.79e-3} & 1.70 & 8.19e-3 & 1.64 \\
11 & 8.12e-3 & 1.18 & 6.24e-3 & 1.54 & \textbf{5.71e-3} & 1.69 \\
13 & 6.44e-3 & 1.16 & 4.46e-3 & 1.62 & \textbf{3.90e-3} & 1.85 \\
15 & 4.30e-3 & 1.19 & 3.01e-3 & 1.75 & \textbf{2.90e-3} & 1.84 \\
17 & 3.27e-3 & 1.13 & \textbf{2.03e-3} & 1.82 & 2.07e-3 & 1.71 \\
19 & 2.69e-3 & 1.07 & 1.37e-3 & 1.98 & \textbf{1.34e-3} & 1.89 \\
21 & 1.64e-3 & 1.22 & \textbf{7.20e-4} & 2.37 & 9.40e-4 & 2.04 \\
23 & 1.10e-3 & 1.06 & 6.20e-4 & 2.18 & \textbf{5.90e-4} & 1.80 \\
\bottomrule
\end{tabular}
\end{table}

As shown in Figure~\ref{fig:compare_selected} and Table~\ref{tab:ler_comparison}, \textsc{QuantiSpect} and \textsc{Accurate} achieve similar LER across most code distances.
Both clearly outperform \textsc{Fast} at moderate and large code distances.
The ordering can be written compactly as
\begin{equation}
\text{LER}_{\text{QS}} \approx \text{LER}_{\text{Accurate}} < \text{LER}_{\text{Fast}} \quad (d \geq 9),
\end{equation}
while \textsc{QuantiSpect} achieves this with $\sim2.71\times$ fewer parameters and $\sim2.84\times$ fewer MACs per voxel than \textsc{Accurate}, shown in Tables~\ref{tab:param_detail} and~\ref{tab:mac}.

We attribute the gap between \textsc{Fast} and the other two methods to the difference in receptive field.
\textsc{Fast} has $R = 9$, while \textsc{Accurate} and \textsc{QuantiSpect} both have $R = 13$.
At small code distances such as $d = 5$ or $7$, the syndrome volume is compact.
An $R = 9$ receptive field already covers most of the relevant local error patterns, so all three methods perform comparably.
As $d$ grows, longer-range local correlations become important.
The larger receptive field of \textsc{Accurate} and \textsc{QuantiSpect} lets them capture patterns that \textsc{Fast} cannot reach.
The reduction factor $\rho$ reflects this.
\textsc{Fast} stays nearly flat across distances, with $\rho \approx 1.1$--$1.2$.
For \textsc{Accurate} and \textsc{QuantiSpect}, $\rho$ grows overall and reaches $\rho \approx 1.8$--$2.4$ at $d = 21$--$23$.

\textsc{QuantiSpect} matches \textsc{Accurate} across $d \geq 9$.
This confirms that the factorized branch design preserves the effective capacity needed for local error correction, even with far fewer parameters.
At the smallest distance $d = 5$, \textsc{QuantiSpect} shows a noticeable accuracy loss.
Its $\rho = 1.01$, compared with $1.24$ for \textsc{Fast} and $1.46$ for \textsc{Accurate}.
At this size the syndrome volume is only $5 \times 5 \times 5$.
It is too small for the separate spatial and temporal branches to contribute useful structure, so the dense architectures perform better.
This overhead disappears by $d = 7$ and is fully absent for $d \geq 9$.

\subsection{Decoding speedup and syndrome density reduction}
\label{sec:speed}

The pre-decoder removes local error contributions from the syndrome, producing a residual with far fewer non-trivial detection events.
Since the MWPM runtime grows polynomially with the syndrome density~$s$, this reduction accelerates the downstream PyMatching decode.
When the PyMatching time saved exceeds the pre-decoder inference cost, the full hybrid pipeline is faster than PyMatching alone.
All inference experiments use FP32 precision on $4\times$A100 GPUs.
Standard PyTorch inference optimizations are enabled, including just-in-time compilation with kernel fusion and a channels-last memory layout.

Table~\ref{tab:latency} reports the per-round PyMatching decode latency at $p = 0.5\%$.
Here ``per-round'' means the total decode time for one shot (covering all $T$ rounds) divided by $T$.
The ``Raw'' column measures PyMatching on the unprocessed syndrome.
The ``Res'' column measures PyMatching on the residual syndrome after the pre-decoder has removed most detection events.
The speedup factor is the ratio of the two.
It reflects how much the syndrome-density reduction accelerates the PyMatching decode.
The pre-decoder forward pass on GPU is timed separately and is not included here.

\begin{table}[!htbp]
\centering
\caption{Per-round PyMatching decode latency ($\mu$s/round, i.e.\ total decode time divided by $T$) at $p = 0.5\%$.
``Raw'' is PyMatching on the unprocessed syndrome.
``Res'' is PyMatching on the residual syndrome after pre-decoding.
The pre-decoder forward pass on GPU is timed separately and is not included.
$\times$ is the speedup factor (Raw\,/\,Res).}
\label{tab:latency}
\footnotesize
\setlength{\tabcolsep}{3pt}
\begin{tabular}{c ccc ccc ccc}
\toprule
& \multicolumn{3}{c}{\textbf{Fast}} & \multicolumn{3}{c}{\textbf{Accurate}} & \multicolumn{3}{c}{\textbf{QuantiSpect}} \\
\cmidrule(lr){2-4} \cmidrule(lr){5-7} \cmidrule(lr){8-10}
$d$ & Raw & Res & $\times$ & Raw & Res & $\times$ & Raw & Res & $\times$ \\
\midrule
5  & 1.53 & 0.96 & 1.60 & 1.50 & 0.96 & 1.57 & 1.56 & 1.03 & 1.52 \\
7  & 2.27 & 1.14 & 1.98 & 2.19 & 0.98 & 2.23 & 2.35 & 1.00 & 2.35 \\
9  & 3.44 & 1.66 & 2.07 & 3.40 & 1.43 & 2.38 & 3.36 & 1.39 & 2.42 \\
11 & 5.25 & 2.41 & 2.18 & 5.21 & 2.10 & 2.49 & 5.48 & 2.07 & 2.65 \\
13 & 7.84 & 3.48 & 2.25 & 7.48 & 2.96 & 2.53 & 7.68 & 2.86 & 2.68 \\
15 & 10.85 & 4.81 & 2.25 & 10.63 & 4.02 & 2.64 & 10.86 & 3.97 & 2.74 \\
17 & 14.63 & 6.36 & 2.30 & 14.37 & 5.25 & 2.74 & 14.15 & 5.25 & 2.70 \\
19 & 19.45 & 8.30 & 2.34 & 19.24 & 6.79 & 2.83 & 19.73 & 6.78 & 2.91 \\
21 & 24.21 & 10.09 & 2.40 & 25.06 & 8.51 & 2.95 & 25.97 & 8.44 & 3.08 \\
23 & 30.29 & 12.25 & 2.47 & 30.75 & 10.01 & 3.07 & 32.00 & 10.30 & 3.11 \\
\bottomrule
\end{tabular}
\end{table}

\textsc{QuantiSpect} achieves the highest speedup factor of the three models across most code distances.
It clearly outperforms \textsc{Fast}, and the gap widens with distance.
It stays close to \textsc{Accurate} throughout.
The small differences between their reported factors reflect run-to-run variation in the raw PyMatching baseline rather than a real gap.
The speedup grows with $d$ for all three models, reaching $3.11\times$ for \textsc{QuantiSpect} at $d = 23$.
This is consistent with the pre-decoder removing a larger fraction of detection events at larger distances.

These results are notable given the model size.
\textsc{QuantiSpect} uses only $0.663$\,M parameters, $2.71\times$ fewer than \textsc{Accurate}'s $1.80$\,M and $1.37\times$ fewer than \textsc{Fast}'s $0.91$\,M.
Despite its smaller size, \textsc{QuantiSpect} matches or exceeds the syndrome reduction of both baselines at moderate and large distances.
The factorized architecture removes local errors more effectively per parameter.
This makes \textsc{QuantiSpect} both smaller and faster at the PyMatching decode than the dense baselines.

\subsection{Effective code distance}
\label{sec:effective_code_distance}

A decoder operating below threshold should drive the logical error rate down quickly as the code distance grows.
The effective code distance quantifies how fast this happens.
Below threshold ($p < p_{\mathrm{th}}$), the logical error rate follows the empirical power law~\cite{dennis2002topological,fowler2012surface}
\begin{equation}
\label{eq:pl_subthreshold}
P_L(p,d) \;\approx\; A \cdot \left(\frac{p}{p_{\mathrm{th}}}\right)^{\!t_d},
\end{equation}
where the exponent $t_d \equiv d_{\mathrm{eff}}(d)/2$ characterizes the effective code distance.
For an ideal decoder that corrects all errors of weight up to $\lfloor(d{-}1)/2\rfloor$, one has $t_d = d/2$.
When the decoder leaves short error segments uncorrected, $t_d$ falls below this ideal value.

To extract $t_d$, we take the basis-averaged data at each code distance $d$ and keep only the sub-threshold points with $p \le p_{\mathrm{sub}}$ and $P_L > 0$.
Each point uses $N = 50000$ shots per basis.
We then fit a Wilson-weighted linear regression in $\log_{10}$ space,
\begin{equation}
\label{eq:step1}
\log_{10} P_L = c + t_d \,\log_{10} \!\left(\frac{p}{p_{\mathrm{th}}}\right),
\end{equation}
where $p_{\mathrm{th}}$ is the method-specific threshold from Table~\ref{tab:fss_threshold}.
The slope $t_d$ is invariant to the choice of $p_{\mathrm{th}}$, since a different threshold only shifts the $x$-axis without changing the slope.
We then fit the resulting $t_d$ values across all distances with a linear model
\begin{equation}
\label{eq:step2}
t(d) = \alpha\,d + \beta,
\end{equation}
so that $\alpha$ measures how fast $t_d$ grows with $d$.
An ideal decoder corresponds to $\alpha = 0.5$.

\begin{table}[!htbp]
	\centering
	\caption{Per-distance $t_d$ at $p_{\mathrm{sub}} = 0.6\%$ (basis-averaged, after pre-decoding).
		$n$ is the number of sub-threshold data points used in the fit.
		Uncertainties are from the weighted least-squares covariance.}
	\label{tab:deff_step1}
	\small
	\setlength{\tabcolsep}{3pt}
	\begin{tabular}{c cc cc cc}
		\toprule
		& \multicolumn{2}{c}{\textbf{Fast}} & \multicolumn{2}{c}{\textbf{Accurate}} & \multicolumn{2}{c}{\textbf{QuantiSpect}} \\
		\cmidrule(lr){2-3} \cmidrule(lr){4-5} \cmidrule(lr){6-7}
		$d$ & $t_d$ & $n$ & $t_d$ & $n$ & $t_d$ & $n$ \\
		\midrule
		5  & $2.84 \pm 0.02$ & 6 & $2.95 \pm 0.05$ & 6 & $2.78 \pm 0.04$ & 6 \\
		7  & $3.70 \pm 0.10$ & 6 & $3.69 \pm 0.17$ & 5 & $3.77 \pm 0.04$ & 6 \\
		9  & $4.25 \pm 0.07$ & 6 & $4.83 \pm 0.08$ & 5 & $4.92 \pm 0.05$ & 5 \\
		11 & $5.15 \pm 0.09$ & 5 & $5.36 \pm 0.09$ & 5 & $5.64 \pm 0.10$ & 5 \\
		13 & $6.00 \pm 0.08$ & 5 & $6.23 \pm 0.16$ & 5 & $6.51 \pm 0.30$ & 4 \\
		15 & $7.21 \pm 0.04$ & 5 & $7.22 \pm 0.16$ & 5 & $7.39 \pm 0.30$ & 5 \\
		17 & $8.12 \pm 0.14$ & 4 & $8.19 \pm 0.13$ & 4 & $8.10 \pm 0.19$ & 4 \\
		19 & $8.37 \pm 0.53$ & 4 & $9.26 \pm 0.25$ & 4 & $8.74 \pm 0.21$ & 4 \\
		21 & $9.76 \pm 0.49$ & 4 & $10.63 \pm 0.68$ & 3 & $10.10 \pm 0.30$ & 4 \\
		23 & $10.73 \pm 0.10$ & 4 & $10.66 \pm 0.20$ & 4 & $10.72 \pm 0.38$ & 3 \\
		\bottomrule
	\end{tabular}
\end{table}

The choice of $p_{\mathrm{sub}}$ balances two competing requirements.
It must be low enough for every fitted point to sit well below the thresholds in Table~\ref{tab:fss_threshold}, where the power law holds, but high enough to retain a sufficient number of non-zero points for a stable fit.
A cutoff of $0.7\%$ would fail the first, since it sits just under \textsc{Fast}'s threshold of $0.740\%$, too close for the asymptotic scaling to apply.
A smaller cutoff would fail the second, leaving too few non-zero points.
We therefore use $0.6\%$, which keeps 3--6 points per distance.
Tables~\ref{tab:deff_step1} and~\ref{tab:deff_step2} then collect the per-distance $t_d$ values and the linear-fit parameters.

\begin{table}[!htbp]
\centering
\caption{Linear fit $t(d) = \alpha\,d + \beta$.
An ideal decoder gives $\alpha = 0.5$; $\alpha < 0.5$ indicates sub-ideal scaling.}
\label{tab:deff_step2}
\begin{tabular}{l c c c}
\toprule
\textbf{Method} & $\alpha$ & $\beta$ & $R^2$ \\
\midrule
Fast        & $0.433 \pm 0.008$ & $0.66 \pm 0.07$ & 0.992 \\
Accurate    & $0.432 \pm 0.009$ & $0.79 \pm 0.09$ & 0.989 \\
QuantiSpect & $0.456 \pm 0.016$ & $0.60 \pm 0.13$ & 0.992 \\
\bottomrule
\end{tabular}
\end{table}

Across all three methods, $t_d$ rises steadily with $d$, confirming that the logical error rate keeps falling as the code distance grows.
The linear fits capture this trend well, with $R^2 \sim 0.99$.
Their slopes stay close to the ideal $0.5$, at $0.433$ for \textsc{Fast}, $0.432$ for \textsc{Accurate}, and $0.456$ for \textsc{QuantiSpect}.
Of the three, \textsc{QuantiSpect} has the steepest slope, so its effective code distance grows fastest with $d$. This means that, compared with \textsc{Fast} and \textsc{Accurate}, \textsc{QuantiSpect} removes more of the short error segments that would otherwise combine into a logical operator.
The remaining gaps between the methods sit within the fitting uncertainty, and at large $d$ the estimates are limited by the few non-zero points that survive the cutoff.

\section{Receptive field scaling: \textsc{QuantiSpect}-21}
\label{sec:rf_scaling}

The modular block design of \textsc{QuantiSpect} lets us enlarge the receptive field by simply adding more blocks.
To test this, we build \textsc{QuantiSpect}-21 by going from $N=5$ to $N=9$ blocks, which raises the receptive field to $R = 1 + 2 + 9 \times 2 = 21$ and the parameter count to $1.18$\,M.
Every other hyperparameter stays identical to the $R=13$ variant, including the hidden dimension, group count, gate ratio, and dropout.
We train QuantiSpect-21 following the general training procedure described in Sec.~\ref{sec:training_details}, but increase the training code distance and the number of syndrome rounds to $d = d_m = 21$ and reduce the initial learning rate to $1 \times 10^{-4}$. We evaluate it under the same conditions, with $T=d$, $N=50000$ shots per basis per point, and uncorrelated PyMatching.

\paragraph{Threshold improvement.}
We run the same FSS analysis and Wilson weighting as in Sec.~\ref{sec:threshold_curves}.
This gives a circuit-level threshold of $p_{\text{th}} = 0.798 \pm 0.002\%$ for \textsc{QuantiSpect}-21, above the $0.769 \pm 0.002\%$ shared by \textsc{QuantiSpect}-13 and the \textsc{Accurate} baseline.
The larger receptive field therefore captures longer-range error correlations that lift the threshold.
The same ranking holds under an unweighted fit, as Appendix~\ref{sec:appendix_fss_scan} shows.

\paragraph{Logical error rate.}
Table~\ref{tab:ler_rf21} compares the basis-averaged LER at $p = 0.5\%$ for the two variants.
They are similar at small distances, but from $d = 13$ onward \textsc{QuantiSpect}-21 reaches a lower LER and a higher $\rho$, with the gap widening as $d$ grows.
These large-distance figures rest on very few observed failures, so they carry substantial statistical uncertainty and should be read as a trend rather than as precise values.
That trend still matches the expectation that a larger receptive field resolves the longer-range error patterns that matter most at large $d$.

\begin{table}[!htbp]
\centering
\caption{Logical error rate (LER) at $p = 0.5\%$ for \textsc{QuantiSpect}-13 ($R=13$, $0.66$\,M) and \textsc{QuantiSpect}-21 ($R=21$, $1.18$\,M).
The reduction factor $\rho = P_L^{\text{no\,predec}} / P_L^{\text{after\,predec}}$ measures the improvement over MWPM-only decoding.}
\label{tab:ler_rf21}
\begin{tabular}{c cc cc}
\toprule
& \multicolumn{2}{c}{\textbf{QS-13}} & \multicolumn{2}{c}{\textbf{QS-21}} \\
\cmidrule(lr){2-3} \cmidrule(lr){4-5}
$d$ & LER & $\rho$ & LER & $\rho$ \\
\midrule
5  & 2.15e-2 & 1.01 & 2.33e-2 & 0.92 \\
7  & 1.20e-2 & 1.50 & 1.18e-2 & 1.47 \\
9  & 8.19e-3 & 1.64 & 7.26e-3 & 1.83 \\
11 & 5.71e-3 & 1.69 & 4.61e-3 & 2.13 \\
13 & 3.90e-3 & 1.85 & 3.28e-3 & 2.41 \\
15 & 2.90e-3 & 1.84 & 2.33e-3 & 2.11 \\
17 & 2.07e-3 & 1.71 & 1.60e-3 & 2.31 \\
19 & 1.34e-3 & 1.89 & 1.10e-3 & 2.32 \\
21 & 9.40e-4 & 2.04 & 5.60e-4 & 2.98 \\
23 & 5.90e-4 & 1.80 & 4.10e-4 & 2.46 \\
\bottomrule
\end{tabular}
\end{table}

\paragraph{Decoding latency.}
Table~\ref{tab:latency_rf21} reports the per-round PyMatching decode latency of the two variants at $p = 0.5\%$, with the same Raw, Res, and speedup definitions used in Sec.~\ref{sec:speed}.
Both variants accelerate the global decoder, and at the largest distances ($d \geq 17$) \textsc{QuantiSpect}-21 consistently reaches the higher speedup, growing to $3.25\times$ at $d = 23$ against $3.11\times$ for \textsc{QuantiSpect}-13.
The deeper network removes more of the syndrome, so PyMatching runs on a sparser residual and finishes sooner.
This speedup is measured on the matching stage alone and does not include the pre-decoder's own inference, whose cost is higher for \textsc{QuantiSpect}-21 because of its larger multiply-accumulate count.
Whether the larger receptive field pays off end to end therefore depends on the balance between this added inference cost and the matching time it saves, as captured by the standalone and hybrid timing model of Eqs.~\ref{eq:standalone_time}--\ref{eq:hybrid_time}.

\begin{table}[!htbp]
\centering
\caption{Per-round PyMatching decode latency ($\mu$s/round) at $p = 0.5\%$ for \textsc{QuantiSpect}-13 ($R=13$, $0.66$\,M) and \textsc{QuantiSpect}-21 ($R=21$, $1.18$\,M).
Raw is the latency without pre-decoding, Res the latency on the pre-decoded residual, and $\times$ their ratio, the speedup factor.}
\label{tab:latency_rf21}
\begin{tabular}{c ccc ccc}
\toprule
& \multicolumn{3}{c}{\textbf{QS-13}} & \multicolumn{3}{c}{\textbf{QS-21}} \\
\cmidrule(lr){2-4} \cmidrule(lr){5-7}
$d$ & Raw & Res & $\times$ & Raw & Res & $\times$ \\
\midrule
5  & 1.56 & 1.03 & 1.52 & 1.49 & 0.99 & 1.50 \\
7  & 2.35 & 1.00 & 2.35 & 2.53 & 0.99 & 2.56 \\
9  & 3.36 & 1.39 & 2.42 & 3.57 & 1.47 & 2.43 \\
11 & 5.48 & 2.07 & 2.65 & 5.20 & 1.97 & 2.64 \\
13 & 7.68 & 2.86 & 2.68 & 7.55 & 2.63 & 2.87 \\
15 & 10.86 & 3.97 & 2.74 & 10.04 & 3.89 & 2.58 \\
17 & 14.15 & 5.25 & 2.70 & 14.35 & 4.99 & 2.87 \\
19 & 19.73 & 6.78 & 2.91 & 18.74 & 6.00 & 3.12 \\
21 & 25.97 & 8.44 & 3.08 & 23.31 & 7.20 & 3.24 \\
23 & 32.00 & 10.30 & 3.11 & 29.76 & 9.14 & 3.25 \\
\bottomrule
\end{tabular}
\end{table}

\paragraph{Effective code distance.}
We fit $t_d$ versus $d$ by the same procedure as in Sec.~\ref{sec:effective_code_distance}.
This gives a slope of $\alpha = 0.481 \pm 0.012$ for \textsc{QuantiSpect}-21, slightly above the $0.456 \pm 0.016$ of \textsc{QuantiSpect}-13.
Both stay closer to the ideal $0.5$ than the baseline slopes of $0.432$ to $0.433$, which confirms that the factorized design preserves the effective-distance scaling even as the receptive field grows.

\paragraph{Comparison with dense models from Ref.~\cite{chamberland2026fast}.}
Table~\ref{tab:rf_nvidia} compares the \textsc{QuantiSpect} variants with all five dense models from Ref.~\cite{chamberland2026fast} by receptive field and parameter count.
The comparison brings out the core efficiency advantage of the structure-aware factorized design.
A dense layer costs $C^2 k^3$ parameters, which grows quadratically in the channel width.
Doubling the width to 256 filters, as in Models~2 and~5, raises the count to $3.60$ and $7.13$\,M without a proportionate accuracy gain in Ref.~\cite{chamberland2026fast}.
The $R=17$ dense model, Model~3, already needs $k=5$ kernels and $4.22$\,M parameters.
The factorized design instead adds only about $128$\,k parameters per block, and each block widens $R$ by $2$, so its parameter count grows only linearly with the receptive field.

\begin{table}[!htbp]
\centering
\caption{Parameter counts of the \textsc{QuantiSpect} variants and all dense models from Ref.~\cite{chamberland2026fast}, ordered by receptive field.}
\label{tab:rf_nvidia}
\small
\begin{tabular}{@{}llcrr@{}}
\toprule
\textbf{Model} & \textbf{Type} & $k$ & $R$ & \textbf{Params} \\
\midrule
1 (Fast)     & Dense, $128{\times}4$ & 3 &  9 & 0.91\,M \\
2            & Dense, $256{\times}4$ & 3 &  9 & 3.60\,M \\
4 (Accurate) & Dense, $128{\times}6$ & 3 & 13 & 1.80\,M \\
5            & Dense, $256{\times}6$ & 3 & 13 & 7.13\,M \\
3            & Dense, $128{\times}4$ & 5 & 17 & 4.22\,M \\
\midrule
\textbf{QS-13} & \textbf{Factor., 5 blk} & \textbf{3} & \textbf{13} & \textbf{0.66\,M} \\
\textbf{QS-21} & \textbf{Factor., 9 blk} & \textbf{3} & \textbf{21} & \textbf{1.18\,M} \\
\bottomrule
\end{tabular}
\end{table}

\textsc{QuantiSpect}-21 reaches $R=21$ with only $1.18$\,M parameters.
That is $3.6\times$ fewer than the $R=17$ dense model at $4.22$\,M, even though it reaches a larger receptive field.
The saving comes from the structure-aware decomposition.
By building the partial separability of spatial and temporal error patterns into the architecture, the factorized branches avoid the full cross-channel mixing that a dense convolution performs at every voxel.
The per-block MAC cost drops by a similar factor, about $3.6\times$ fewer than a dense layer in Table~\ref{tab:mac}, so the advantage compounds as depth increases.
Under the same counting convention, QuantiSpect-21 requires approximately
\(1.124\,\mathrm{M}\) convolutional MACs per voxel, which is \(1.60\times\) fewer
than that of the \(R=13\) \textsc{Accurate} baseline and approximately \(3.76\times\)
fewer than that of the \(R=17\) dense model.

These results show that the \textsc{QuantiSpect} architecture offers a scalable route to larger receptive fields.
A dense network pays a large parameter cost for each unit of receptive field, whereas the factorized design pays far less.
\textsc{QuantiSpect}-21 reaches a larger receptive field than every dense model in Ref.~\cite{chamberland2026fast}, yet it still uses fewer parameters than the $R=13$ \textsc{Accurate} baseline.

\section{Conclusion and Outlook}
\label{sec:conclusion}

We have presented \textsc{QuantiSpect}, a lightweight 3D CNN pre-decoder for the rotated surface code.
It replaces dense convolutions with three factorized branches that separately capture spatial, temporal, and mixed spatio-temporal error patterns.
The base variant \textsc{QuantiSpect}-13 uses $N=5$ blocks to reach receptive field $R=13$. Compared to the standaalone PyMatching, \textsc{QuantiSpect}-13  lowers the logical error rate by a factor $\rho \approx 1.85$ at $d=13$ and $p=0.5\%$. Compared to the  \textsc{Accurate} baseline, it reaches a circuit-level threshold $p_{\text{th}} \approx 0.77\%$ that matches the \textsc{Accurate} baseline while using $\sim2.71\times$ fewer parameters and $\sim2.84\times$ fewer per-voxel MACs.

The modular block design lets us scale the receptive field by stacking more blocks.
The expanded variant \textsc{QuantiSpect}-21 uses $N=9$ blocks to reach $R=21$. It raises the circuit-level threshold to $p_{\text{th}} \approx 0.80\%$, and lowers the logical error rate a further $31\%$ over \textsc{QuantiSpect}-13 at the codition $d=23$ and $p=0.5\%$, to a factor $\rho = 2.46$ over standalone PyMatching.
Even with a larger receptive field, it uses only $1.18$\,M parameters, $3.6\times$ fewer than the $4.22$\,M $R=17$ dense model of Ref.~\cite{chamberland2026fast}.

Across $d = 5$ to $23$, \textsc{QuantiSpect}-13 matches the \textsc{Accurate} baseline for $d \geq 9$ and beats the \textsc{Fast} baseline at moderate and large distances, and \textsc{QuantiSpect}-21 improves further for $d \geq 13$.
This ordering tracks the receptive field, since \textsc{QuantiSpect}-13 and \textsc{Accurate} share $R = 13$ and perform alike, while the gains over the $R = 9$ \textsc{Fast} model and from the $R = 21$ variant come from a larger field.
A larger field captures longer-range error correlations, which matter more as $d$ grows, so the gaps widen at large distances.

A sub-threshold scaling analysis fits $t_d \sim \alpha\,d$ and gives slopes $\alpha \approx 0.43$--$0.48$ for all models, close to the ideal $0.5$, with $R^2 \sim 0.99$.
The two \textsc{QuantiSpect} variants sit closest to ideal, at $\alpha = 0.456$ and $0.481$ against $0.432$--$0.433$ for the baselines.
The factorized design therefore preserves most of the error-correcting power of the surface code.

Physical errors and their detection events are partially separable in space and time, and building this structure into the network makes the receptive field cheap to enlarge.
Each added block raises $R$ by $2$ at about $128$\,k parameters.
A dense $k=3$ layer adds the same $2$ but costs $442$\,k, so a dense network scaled by depth pays $3.4\times$ more per stage.
Scaling a dense network by kernel size instead, as in Model~3 ($R=17$, $k=5$), makes the per-layer cost grow as $k^3$ and, at fixed depth, cubically in $R$.
Our design keeps the kernel fixed and adds blocks, so its cost grows only linearly in $R$.
As quantum error correction moves toward real-time operation at large distances, this efficiency matters more and more.

Several steps would carry \textsc{QuantiSpect} from memory experiments toward a practical decoder.
The most immediate is to train on noise calibrated to real hardware rather than the idealized circuit-level model used here, so the pre-decoder adapts to a specific device.
The architecture should then move beyond memory to logical operations, starting with lattice surgery~\cite{horsman2012surface} and reaching toward the non-Clifford operations needed for universal computation.
A lightweight pre-decoder is especially useful there, since merged patches reach large effective distances and carry the highest matching load.
Together these steps would turn the efficiency shown here into a full fault-tolerant decoding stack.

The model and code are publicly available at \url{https://huggingface.co/quantispect/QuantiSpect-V1}.

\appendix
\section{Per-distance model comparisons}
\label{sec:appendix_compare}

\begin{figure}[!ht]
\centering
\includegraphics[width=\columnwidth]{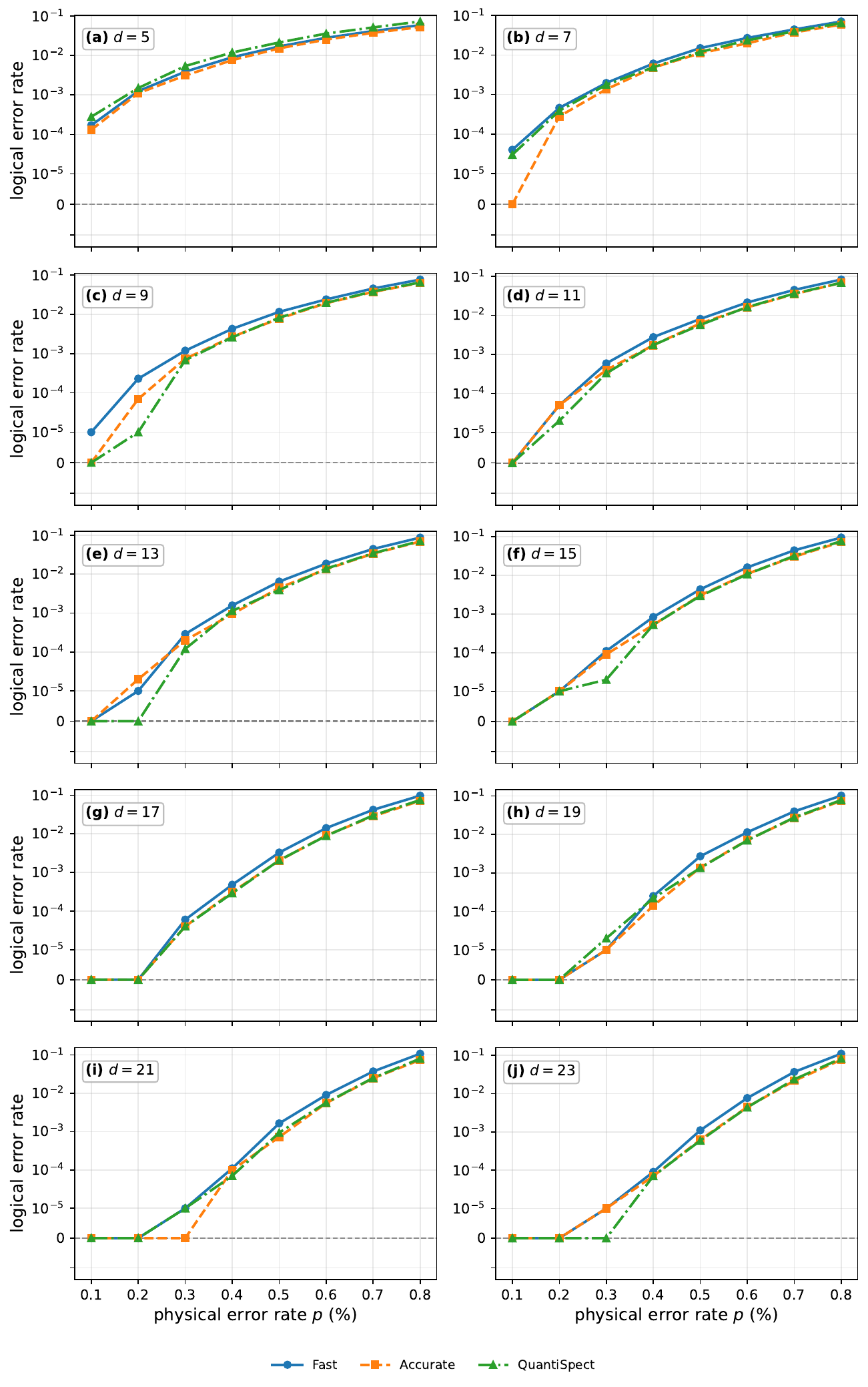}
\caption{Head-to-head logical error rate comparison across all ten code distances, panels (a)--(j) for $d = 5$ to $23$ (each with $T = d$ measurement rounds, $N = 50000$ shots per basis per point).
Methods compared: \textsc{Fast} (blue, solid), \textsc{Accurate} (orange, dashed), and \textsc{QuantiSpect} (green, dash-dotted).
The vertical axis uses a symmetric-log scale; the dashed line marks the $\text{LER} = 0$ baseline.}
\label{fig:compare_all}
\end{figure}

Figure~\ref{fig:compare_all} presents the complete set of head-to-head comparison plots for all ten code distances evaluated, supplementing the representative subset shown in Fig.~\ref{fig:compare_selected}.

\section{FSS fitting: weighting comparison and robustness scan}
\label{sec:appendix_fss_scan}

Sec.~\ref{sec:threshold_curves} reports Wilson-weighted finite-size scaling thresholds as the primary result.
This appendix compares that choice against an unweighted fit and an event-count-based robustness check, then details the 80-configuration scan referenced in the main text.

\subsection{Unweighted vs.\ Wilson-weighted fits}
\label{sec:appendix_fss_weighting}

Table~\ref{tab:fss_threshold_full} repeats the FSS fit for every method under both weighting schemes, including \textsc{QuantiSpect}-21 (Sec.~\ref{sec:rf_scaling}).

\begin{table}[!htbp]
\centering
\caption{Circuit-level threshold estimates $p_{\text{th}}$ from finite-size scaling fits over the near-threshold window $p \in [0.3\%, 0.8\%]$ with $d \geq 7$, under both weighting schemes.
``Unwt'' uses raw LER values; ``Wilson'' incorporates Wilson-score 95\,\% CI from $N = 50000$ shots.
Uncertainties are from the curve-fit covariance and do not include systematic fitting-window effects.
RMSE is the collapse residual in $\log_{10}$ space.}
\label{tab:fss_threshold_full}
\begin{tabular}{l l c c c}
\toprule
\textbf{Method} & \textbf{Fit} & $p_{\text{th}}$ (\%) & $\nu$ & RMSE \\
\midrule
Fast        & Unwt   & $0.678 \pm 0.015$ & $0.89$ & $0.16$ \\
            & Wilson & $0.740 \pm 0.002$ & $1.29$ & $0.43$ \\
\addlinespace
Accurate    & Unwt   & $0.737 \pm 0.018$ & $1.13$ & $0.14$ \\
            & Wilson & $0.769 \pm 0.002$ & $1.26$ & $0.33$ \\
\addlinespace
QuantiSpect-13 & Unwt   & $0.733 \pm 0.020$ & $1.11$ & $0.17$ \\
            & Wilson & $0.769 \pm 0.002$ & $1.25$ & $0.36$ \\
\addlinespace
QuantiSpect-21 & Unwt   & $0.748 \pm 0.020$ & $1.11$ & $0.16$ \\
            & Wilson & $0.798 \pm 0.002$ & $1.23$ & $0.30$ \\
\bottomrule
\end{tabular}
\end{table}

Under the unweighted fit, every method yields a lower threshold than under Wilson weighting, by $0.03$--$0.06$ percentage points, but the collapse RMSE is roughly $2\times$ tighter ($0.14$--$0.17$ vs.\ $0.30$--$0.43$ in $\log_{10}$ space).
This gap occurs because Wilson weighting gives much more weight to near-threshold points, where the relative uncertainty in $\log_{10}$ space is smallest, and correspondingly less weight to the more numerous low-$p$ points.
This raises $p_{\text{th}}$ but worsens the fit to the bulk of the data.
The qualitative ranking of methods, \textsc{QuantiSpect}-21 $>$ \textsc{QuantiSpect}-13 $\approx$ \textsc{Accurate} $>$ \textsc{Fast}, is identical under both weighting schemes, so this choice does not affect the paper's conclusions.

\begin{figure*}[t]
\centering
\includegraphics[width=\textwidth]{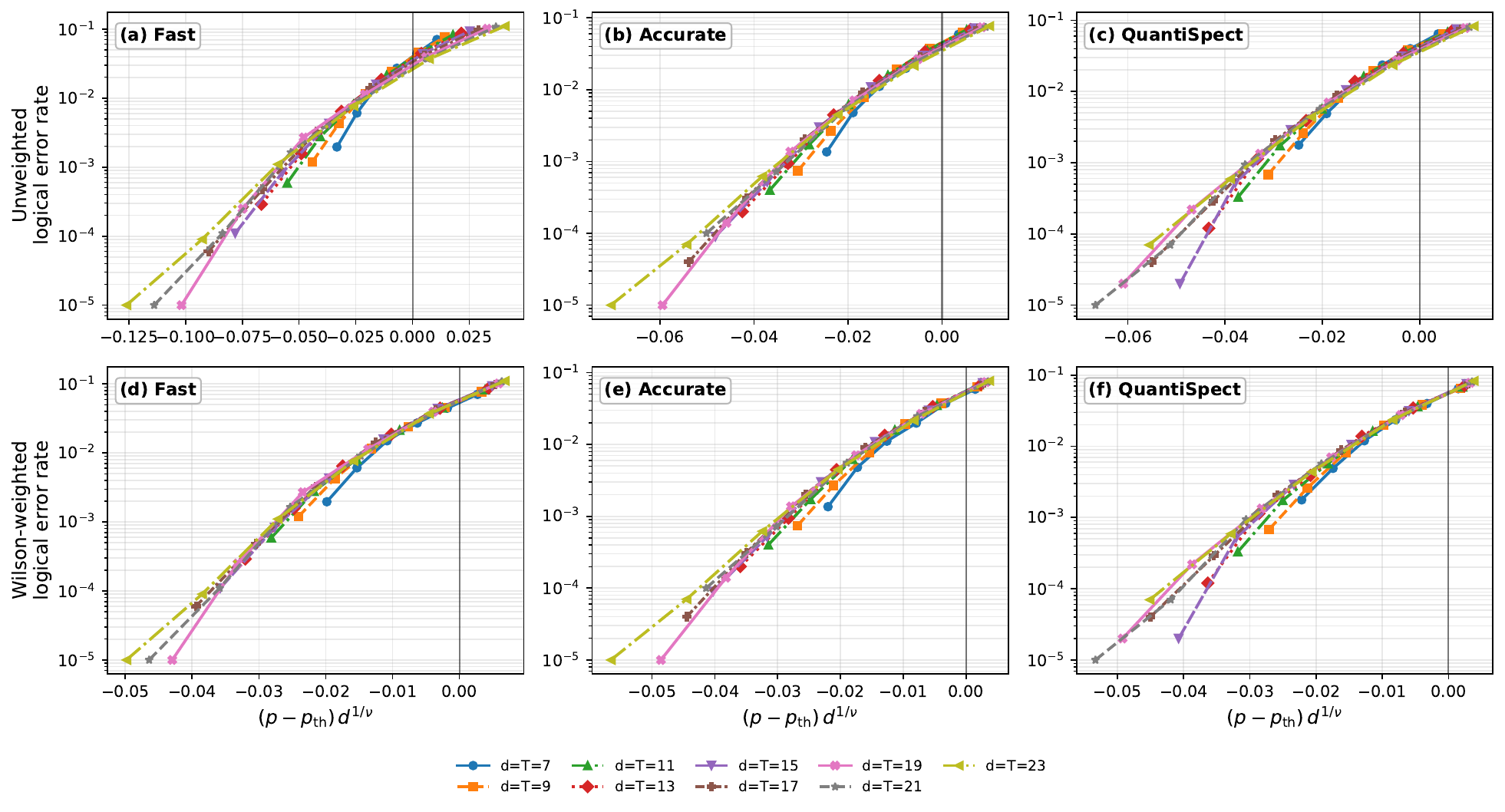}
\caption{Finite-size scaling data collapse for the three pre-decoder methods.
Panels (a)--(c) use the unweighted fit; panels (d)--(f) use the Wilson-weighted fit.
Columns are \textsc{Fast} (a,\,d), \textsc{Accurate} (b,\,e), and \textsc{QuantiSpect}-13 (c,\,f).
Both fits use a second-order polynomial ansatz over the window $p \in [0.3\%,\, 0.8\%]$ with $d \geq 7$ ($9$ distances, $\sim54$ points per method after excluding zero-LER entries).
The horizontal axis is the rescaled variable $(p - p_{\text{th}})\, d^{1/\nu}$, the vertical axis is the logical error rate on a logarithmic scale.
The fitted thresholds $p_{\text{th}}$ (unweighted\,/\,Wilson-weighted) are \textsc{Fast} $0.678\%$ / $0.740\%$, \textsc{Accurate} $0.737\%$ / $0.769\%$, and \textsc{QuantiSpect}-13 $0.733\%$ / $0.769\%$.}
\label{fig:fss_collapse_full}
\end{figure*}

Figure~\ref{fig:fss_collapse_full} shows the FSS collapse under both weighting schemes for \textsc{Fast}, \textsc{Accurate}, and \textsc{QuantiSpect}-13.

\subsection{Event-count-based robustness check}
\label{sec:appendix_fss_mincount}

As a further check, we also refit each method with an unweighted least-squares fit after excluding points with fewer than a minimum number of estimated observed logical-failure events, following the event-count filtering approach used in some surface-code threshold studies~\cite{stephens2014fault} as an alternative to explicit statistical weighting.
As the event-count cutoff tightens from $0$ (no filtering) to $2000$, $p_{\text{th}}$ generally rises from the unweighted value toward the Wilson-weighted regime for every method, though not strictly monotonically.
For example, \textsc{QuantiSpect}-21 moves from $0.748\%$ (no filter) through $0.766\%$ (a cutoff of $100$ events) to $0.798\%$ (Wilson-weighted).
The ranking between methods is preserved at every cutoff, and the gap between \textsc{QuantiSpect}-21 and the baselines widens rather than narrows as the cutoff tightens, corroborating that this gap is not an artifact of a small number of noisy, low-count points.
This check is exploratory and not part of the reported fitting protocol.
It is included only as supporting evidence for the robustness of the Wilson-weighted results reported in the main text.

\subsection{Robustness scan configurations}
\label{sec:appendix_fss_scan_configs}

The 80 fitting configurations used in the robustness scan of Sec.~\ref{sec:threshold_curves} are generated as the Cartesian product of the following parameter axes:

\begin{itemize}[nosep,leftmargin=1.5em]
    \item \textbf{Polynomial order} of $f$: $\{2,\;3\}$.
    \item \textbf{Error-rate window} $[p_{\min},\,p_{\max}]$: full range, $(0.2\%,\,0.7\%)$, $(0.3\%,\,0.7\%)$, $(0.3\%,\,0.8\%)$, $(0.2\%,\,0.8\%)$.
    \item \textbf{Minimum distance} $d_{\min}$: all $(d \geq 5)$, $7$, $9$, $11$.
    \item \textbf{Weighting}: unweighted, or Wilson-score 95\,\% CI weighted.
\end{itemize}

\noindent This yields $2 \times 5 \times 4 \times 2 = 80$ configurations per pre-decoder method.
For each configuration, the FSS ansatz $\log_{10}P_L = f\!\bigl((p - p_{\text{th}})\,d^{1/\nu}\bigr)$ is fitted via nonlinear least squares (\texttt{scipy.optimize.curve\_fit}), yielding a threshold estimate $p_{\text{th}}$, critical exponent $\nu$, and collapse RMSE in $\log_{10}$ space.
The ranges reported in Sec.~\ref{sec:threshold_curves} ($0.55\%$--$0.74\%$ for \textsc{Fast}, $0.67\%$--$0.78\%$ for \textsc{Accurate}, $0.60\%$--$0.78\%$ for \textsc{QuantiSpect}) correspond to the minimum and maximum $p_{\text{th}}$ across all 80 configurations for each method.

\begin{acknowledgments}
P.~G. thanks to the National Natural Science Foundation of China (Grant No.~62501058).
J.-Z.~H. thanks to the Natural Science Foundation of Jiangsu Province (No.~BK20250404) and the Youth Science and Technology Talent Support Program of Jiangsu Province (No.~JSTJ-2025-600) and Suzhou (No.~2025(062)).
X.-D.~L. thanks to the National Key R\&D Program of China (No.~2025YFF0515500) and the Shanghai Municipal Science and Technology Commission Strategic Frontier Special Project (No.~25DP2600100).
G.-L.~L. thanks to the National Natural Science Foundation of China (Grant No.~62131002).
\end{acknowledgments}

\bibliographystyle{unsrt}
\bibliography{refs}

\end{document}